\documentclass[fleqn,usenatbib]{mnras}

\usepackage{newtxtext,newtxmath}

\usepackage[T1]{fontenc}
\usepackage{ae,aecompl}

\usepackage{graphicx}	

\usepackage{amsmath}	
\usepackage{amssymb}	

\title[Radio Study of PWN CTB\,87]{A Radio Continuum and Polarisation Study of the pulsar wind nebula CTB\,87 (G74.9+1.2)}

\author[R. Kothes, W. Reich, S. Safi-Harb, B. Guest, P. Reich, E. F\"{u}rst]{
R. Kothes,$^{1}$
W. Reich,$^{2}$
S. Safi-Harb,$^{3}$
B. Guest,$^{3}$
P. Reich,$^{2}$
E. F\"urst$^{2}$
\\
$^{1}$ National Research Council Canada, Herzberg Astronomy and Astrophysics, 
Dominion Radio Astrophysical Observatory,\\ P.O. Box 248, Penticton, B.C., V2A 6J9, Canada\\
$^{2}$ Max-Planck-Institut f\"ur Radioastronomie, Auf dem H\"ugel 69, D-53121, Bonn, Germany\\
$^{3}$Department of Physics and Astronomy, University of Manitoba, Winnipeg, MB, R3T 2N2, Canada\\}

\date{Accepted XXX. Received YYY; in original form ZZZ}

\pubyear{2019}

\begin{document}
\label{firstpage}
\pagerange{\pageref{firstpage}--\pageref{lastpage}}
\maketitle

\begin{abstract}
We present radio continuum and linear polarisation observations of the pulsar wind nebula CTB\,87 (G\,74.9$+$1.2) with the Effelsberg 100-m radio telescope between 4.75 and 32~GHz. An analysis of these new data including archived low-frequency observations at 1420~MHz and 408~MHz from the Canadian Galactic Plane Survey
shows that CTB\,87 consists of two distinct emission components: a compact kidney-shaped component, 14~pc $\times$ 8.5~pc ($7\farcm8 \times 4\farcm8$) in size and a larger diffuse, spherical and centrally peaked component of about 30~pc ($17\arcmin$) in diameter. The kidney-shaped component with a much steeper radio continuum spectrum is highly linearly polarised and likely represents a relic pulsar wind nebula. The diffuse component represents the undisturbed part of the PWN expanding inside a cavity or stellar wind bubble. The previously reported spectral break above 10~GHz is likely the result of missing large-scale emission and insufficient sensitivity of the high-frequency radio continuum observations. The simulation of the system's evolution yields an age of about 18,000 years as the result of a type II supernova explosion with an ejecta mass of about 12~M$_\odot$ and an explosion energy of about $7\times 10^{50}$~erg. We also found evidence for a radio shell in our polarisation data which represents the blast wave that entered the molecular cloud complex at a radius of about 13~pc.
\end{abstract}




\section{Introduction}

The pulsar wind nebula (PWN) CTB\,87 (G\,74.9+1.2) is one member of a small but
growing class of supernova remnants (SNRs)
showing a filled-centre radio morphology (also historically referred to as plerions or Crab-like SNRs), 
in contrast to the more common
shell-type SNRs with the brightest emission coming from an arc-like
structure at the rim of the source. The PWNe in general have flatter radio 
synchrotron spectra than the shell-type SNRs. 
Their morphology is believed to be
the result of a synchrotron-emitting nebula powered by a fast rotating
neutron star, which releases a highly relativistic magnetised particle wind. 

CTB\,87 is a member of this class of pure filled-centre PWNe with no detected shell. A recent \textit{XMM-Newton} study aimed at searching for the SNR shell and the supernova ejecta did not detect any thermal X-ray emission from the SNR \citep{gues19}.
The radio spectral index
of CTB\,87 is $\alpha = -0.29 \pm 0.02$ ($S~\propto~\nu^\alpha$) up to a frequency
of about 10~GHz \citep{koth06}. Above 11~GHz the spectrum
seems to have a break and the spectral index changes to 
$\alpha \approx -1.08$ \citep{mors87}. The origin of this proposed break in the radio spectrum
is not understood.

\begin{table*}
\caption{\label{tab:param} Parameters of the receiver systems and the mapping procedure for the
Effelsberg Observations (FET = Field-Effect Transistor, HEMT = High-Electron-Mobility Transistor)}
\centerline{\begin{tabular}{lcccc}
\hline
\noalign{\smallskip}
Frequency [GHz]           & 4.75   & 10.55   & 14.7    & 32.0 \\
\noalign{\smallskip}
\hline
\noalign{\smallskip}
Feeds                     & 2      & 4       & 4       & 3  \\
Receivers                 & 3~$\times$~FET   & 8~$\times$~HEMT & 4~$\times$~HEMT
     & 6~$\times$~Mixer/6~$\times$~HEMT   \\
T$_\mathrm{sys}$ [K]     & 60      & 50      & 200     & 500/130 \\
Bandwidth [MHz]          & 500     & 300     & 1000    & 2000 \\
HPBW [\arcmin ]          & 2.5     & 1.2     & 0.85    & 0.45 \\
Scanning velocity [\arcmin /min] & 60    & 40    & 25      & 20 \\
Step interval [\arcsec ]         & 60    & 20    & 15      & 10 \\
Observed field size [$\arcmin \times \arcmin$] & 28$\times$28 & 31$\times$15 &
   16$\times$16 & \\
Map size [$\arcmin \times \arcmin$]  & 20$\times$20 & 15$\times$15 &
   15$\times$15 & 10$\times$10 \\
Date of observation  & 10./12.1989 & 10.1994 & 01.1996 & 03.1989/12.1996 \\
Calibrator for total intensity & 3C286  & 3C286 & NGC7027 & 3C286\\
Flux density [Jy]        & 7.5 & 4.5  & 5.75 & 2.1  \\
Calibrator for polarised intensity & 3C286  & 3C286 & --- & Cygnus A West \\
Linear polarisation [\%]  & 11.3 & 11.8 & --- & 9.6 \\
Polarisation angle [\degr ] & 33 & 33 & --- & 133 \\
Number of coverages      & 4 & 2 & 1 & 14 \\
r.m.s. total intensity [mJy/beam] & 2 & 3 & 10 & 10 \\
r.m.s. polarisation [mJy/beam] & 1 & 0.8 & --- & 2 \\
Conversion Factor T$_b$/S [K/(Jy/beam)] & 2.41 & 2.12 & 2.17 & 1.64 \\
\noalign{\smallskip}
\hline
\end{tabular}}
\end{table*}

\citet{math13} presented 
an 87~ks \textit{CHANDRA} high-resolution X-ray study of CTB\,87. They detected the point-like source CXOU~J201609.2+371110, the putative neutron
star that powers CTB\,87, surrounded by a compact ($<$10$^{\prime\prime}$-radius) nebula whose morphology
suggests the presence of a torus-like structure and a possible jet. They also found a diffuse nebula extending
from the point source to the north-west (in equatorial coordinates). A 
spectral analysis of the three X-ray components
indicates that CTB\,87 is an evolved PWN, which already interacted with the
reverse shock of the supernova explosion. The kidney-shaped nebula is likely
a ``relic'' PWN similar in its evolutionary phase to Vela-X and G327.1$-$1.1 \citep{math13,gues19}. 

The crowded Cygnus~X 
region around CTB\,87 and the compact nature of the X-ray nebula make it very challenging to resolve this PWN in $\gamma$-rays. A first detection of $\gamma$-rays from this direction of the sky by the MILAGRO $\gamma$-ray observatory \citep[MGRO~J2019+37,][]{abdo07} was later resolved by VERITAS  (Very Energetic Radiation Imaging Telescope Array System) with VER~J2016+371 being spatially coincident with CTB\,87 \citep{aliu14}. 
{\it Fermi}-LAT found a source nearby,
which \citet{abey18} argued to be the result of two unresolved high-energy $\gamma$-ray sources whose position and variability depend on the energy used.
When the {\it Fermi} emission is modelled as two point sources, a single power-law is required to fit the CTB\,87 point source and the TeV VERITAS emission. The overlap between the TeV emission from VER~J2016+371 and the peak of radio emission from CTB\,87 supports their association and the evolved nature of the PWN \citep{aliu14}.

The distance to CTB\,87 
was estimated based on HI-absorption measurements and CO observations
of possibly related molecular material to be $d~\approx~6.1$~kpc
\citep{koth03}. A more detailed study of the nearby molecular cloud complex by \citet{liu18} 
confirmed this. \citet{liu18} also proposed that CTB\,87 is located at an off-centre position inside 
a large cavity or stellar wind bubble in the interstellar medium (ISM), which is also inhabited by 82 OB
star candidates.

We present radio continuum observations of CTB\,87 at 4.75~GHz, 10.55~GHz, 14.7~GHz,
and 32~GHz made with the Effelsberg 100-m radio telescope 
including linear polarisation at 4.75~GHz, 10.55~GHz, and 32~GHz.
We also analysed archival 408-MHz
and 1420-MHz data from the Canadian Galactic Plane 
Survey \citep[CGPS][]{tayl03}, including the linearly-polarised
emission at 1420~MHz \citep{land10}.

In Section~2, we describe the
observations and data-processing procedures of the Effelsberg and DRAO Synthesis Telescope
data and present the results in Section~3. A thorough analysis of the 
results and their interpretation is presented in Section 4. 

\section{Radio continuum observations}

\subsection{Effelsberg Observations}

The Effelsberg 100-m telescope was used to map CTB\,87 at 4.75~GHz,
10.55~GHz, 14.7~GHz, and 32~GHz. At all frequencies, two-feed or
multiple-feed systems installed in the secondary focus of the telescope
were used. At 4.75~GHz, 10.55~GHz, and 32~GHz the two circularly-polarised wave components from each feed were recorded as a measure of the total intensity and via IF-correlation simultaneously the linearly-polarised emission. At 14.7~GHz, one circularly-polarised channel per feed was recorded to measure total intensities. The 4.75-GHz receiver
is a correlation system, while all others are highly stable total-intensity receivers. With the exception of the early 32-GHz mixer system, all
receivers are cooled. Calibrators were observed several times during each observing run. Fluctuations among them are typically less than 3\%. More observational parameters are listed in Table~\ref{tab:param}.

All observations were made 
by continuously tracking the source position centre and mapping the surrounding area by scanning along the azimuth direction. The standard data reduction, which is based on the NOD2 format \citep{hasl74}, was applied. The multi-feed observations at 4.75~GHz, 10.55~GHz, and 32~GHz were restored using the standard algorithm described by \citet{emer79}.
At 14.7~GHz, the \citet{emer79} algorithm cannot be used, since it requires a multi-beam system with feeds aligned along azimuth. The 14.7~GHz system had four feed in a rotatable square. This observation was restored using the program DBMEM \citep{rich92}, based on the MEM algorithm (MEM = Maximum Entropy Method), which was adapted to process Effelsberg multi-beam observations by \citet{koth94}. At
10.55~GHz, the sidelobes of the nearby compact source E\,2013+3702 were
removed using the CLEAN program written for Effelsberg observations 
by \citet{klei95}. At 14.7~GHz, the bright compact source
E\,2013+3702 was removed by subtracting a beam
pattern observed on the source 3C84. Baseline improvements by unsharp
masking \citep{sofu79} and the `PLAIT' algorithm
described by \citet{emer88} were
applied to all observations. 

\begin{table}
\caption{CGPS Survey characteristics for the area around CTB\,87.}
\label{obspara}
\begin{center}
\begin{tabular}{|lrl|}
\hline
\multicolumn{2}{|l} {Operating frequencies:} & 1420\,MHz \\
                            &              & 408\,MHz \\
Angular resolution:         & 1420\,MHz    & $49\arcsec \times 1\arcmin 12\arcsec$ \\
                            & 408\,MHz     & $2\farcm 8 \times 4\farcm 4$ \\
Continuum sensitivity:      & 1420\,MHz    & 0.3 mJy/beam rms \\
                            & 408\,MHz     & 6 mJy/beam rms \\
Conversion Factor & 1420\,MHz & 0.1718\\
$T_{\rm b}/S$ [K/(mJy/beam)] & 408\,MHz  & 0.1655 \\
\hline
\end{tabular}
\end{center}
\end{table}

\begin{figure*}
\centerline{\includegraphics[height=0.95\textwidth,angle=-90,clip]{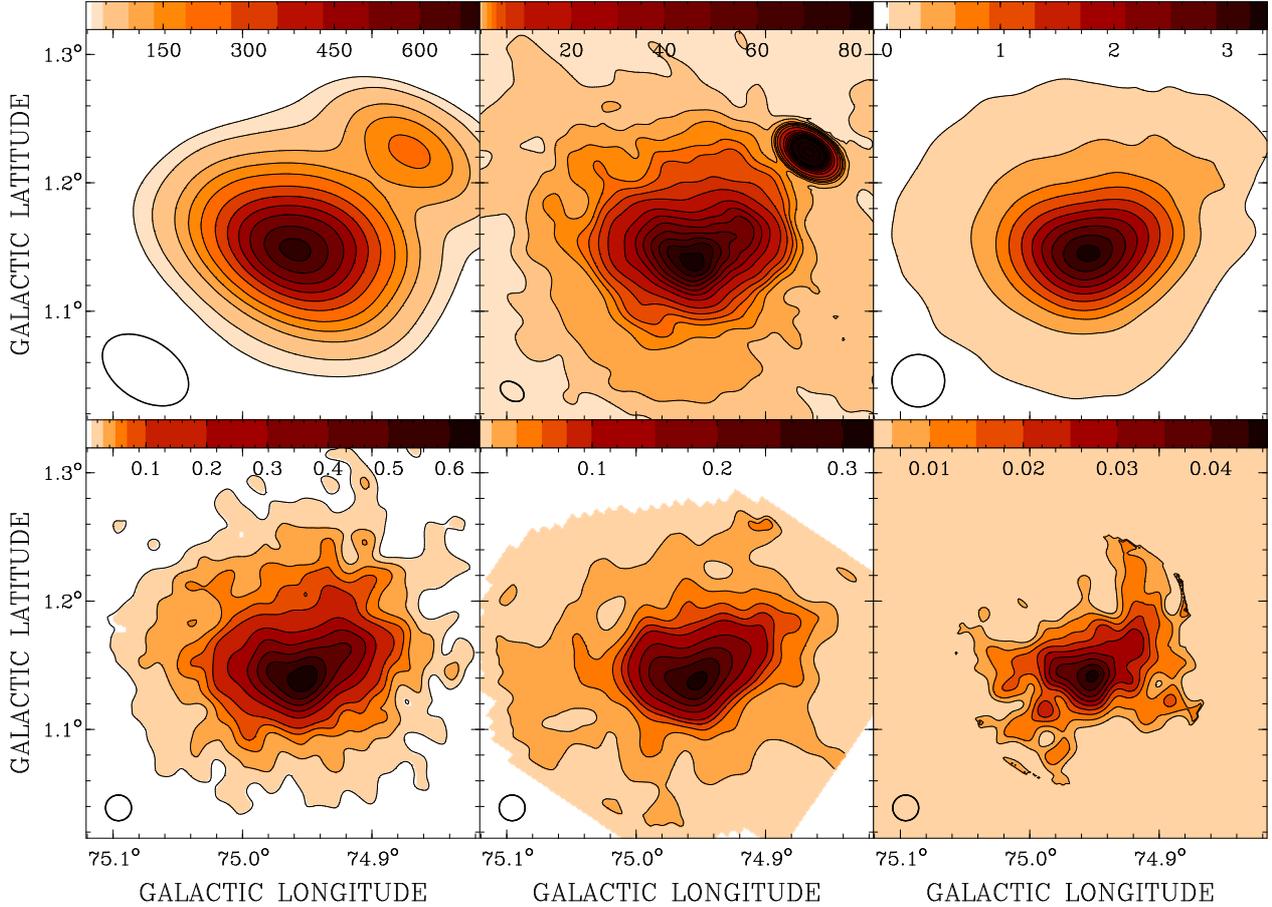}}
\caption{Total-power images of CTB\,87 at 408~MHz (top left), 1420~MHz (top centre), 4.75~GHz (top right), 10.55~GHz (bottom left), 14.7~GHz (bottom centre), and 32~GHz (bottom right). Map units are brightness temperature in K. The black contours indicate the steps for the shading pattern. Their levels are indicated by the steps in the colour bars at the top of each image. The resolution of each map is indicated in the lower left corner of each image. The Effelsberg maps at 14.7~GHz and 32~GHz have been convolved to 1\farcm 2.  The nearby compact source E2013+3702 can be seen to the top right of CTB\,87 in the 408~MHz and 1420~MHz images.}
\label{fig:tp}
\end{figure*}

\subsection{DRAO Observations}

The Canadian Galactic Plane Survey (CGPS) is described in detail in \citet{tayl03}. 
For the research presented here, we used those 
parts of the CGPS database which derive from observations with the 
DRAO Synthesis Telescope \citep[DRAO ST,][]{land00}. 
The characteristics of the survey relevant to
the data presented here are listed in Table \ref{obspara}. The survey area is
covered by observations of individual fields whose centres lie on a
hexagonal grid of spacing $112\arcmin$. Angular resolution varies as
cosec(declination) and therefore slowly changes across the final mosaics.
The values given in Table~\ref{obspara} are characteristics at the centre
of CTB\,87. The sensitivity
is limited by confusion due to complex extended structure in the Cygnus
region, not by thermal noise. Before assembly into a mosaic, the data for the
individual fields were carefully processed to remove artefacts and to
obtain the highest dynamic range, using the routines described by
\citet{will99}.

Accurate representation of all structures to the largest scales is
assured by incorporating data from large single-dish antennas with data
from the Synthesis Telescope, after suitable filtering in the Fourier
domain. Procedures are described in \citet{tung17} for 408~MHz, \citet{tayl03} for total power at 1420~MHz and \citet{land10} for the 1420-MHz linearly polarised component. The single-antenna observations were drawn from the 408-MHz All-Sky survey of \citet{hasl82}, the 1.4-GHz Effelsberg survey for 1420-MHz total power \citep{reic90} and the Effelsberg Medium Latitude Survey (EMLS) for the linearly polarised emission at 1420~MHz \citep{reic04}. 

\section{Results}

\subsection{CTB\,87 in radio continuum}

Total-power images of the new radio continuum observations carried out with the Effelsberg telescope along with images from the DRAO ST are presented in Fig.~\ref{fig:tp}. The measurements at 14.7~GHz and 32~GHz have been convolved to 1\farcm2
to increase the signal-to-noise ratio. The bright compact source E\,2013+3702 has
been subtracted from all Effelsberg maps, but we kept E\,2013+3702 in the
displayed CGPS maps at 408 and 1420~MHz to indicate that its
proximity may cause interference. The integrated total-flux
densities of CTB\,87 are
summarised in Table~\ref{tab:flux}. The radio flux densities have been integrated 
in concentric rings of 0\farcm4 width starting at the radio peak of CTB\,87 
at 1420~MHz ($\mathrm{\alpha_{J2000}=
20^h 16^m 11.3^s,\  \delta_{J2000}}=37\degr 11\arcmin 45\arcsec$) up to the first minimum in the radial profile.

\begin{table}
    \centering
    \begin{tabular}{rrcccc}
    \hline
    $\nu$ [MHz] & $S$ [Jy] & $S_c$ [Jy] & $PI$ [Jy] & $\%$-pol. & peak-pol.\\
    \hline
     74 & 14.5$\pm$3.0 & --- &--- & --- & --- \\
     408 & 11.9$\pm$0.9 & --- &--- & ---  & ---\\
1420 & 9.0$\pm$0.8 & 4.6$\pm$0.5 & 0.12$\pm$0.01 & 1.3$\pm$0.2 & 6\,\% \\
4750  & 5.7$\pm$0.4 & --- & 0.46$\pm$0.04 &8.1$\pm$0.9 & 16\,\% \\
10550 & 4.4$\pm$0.3 & 2.0$\pm$0.2 & 0.35$\pm$0.06 & 8.0$\pm$1.5 & 17\,\% \\
14700 & 2.9$\pm$0.5 & 1.7$\pm$0.2 & --- & --- & --- \\
32000 & 1.9$\pm$0.4 & 1.3$\pm$0.3 & 0.14$\pm$0.02 & 7.4$\pm$1.9 & 19\,\% \\
\hline
    \end{tabular}
    \caption{Integrated total-flux densities $S$, flux
densities of the compact component $S_c$, and
polarised intensities $PI$ of CTB\,87 at frequencies $\nu$. The 74-MHz
flux density was integrated from a map taken from the VLSSr \citep{vlss}.
The 408-MHz flux density was taken from the CGPS SNR catalogue
\citep{koth06}.}
    \label{tab:flux}
\end{table}

All total-power observations reveal the well-known kidney-shaped structure
of CTB\,87. In addition, an underlying extended diffuse component is seen,
which is most pronounced in the 10.55-GHz and 1420-MHz maps. Just a visual inspection of the images already indicates that this 
diffuse component has a somewhat flatter spectrum than the kidney-shaped 
feature, because it seems to become more prominent with increasing frequency, with the exception of the 32-GHz map. We will
discuss the two components in Sect. 3.2.

\begin{figure}
\centerline{\includegraphics[width=0.47\textwidth,clip]{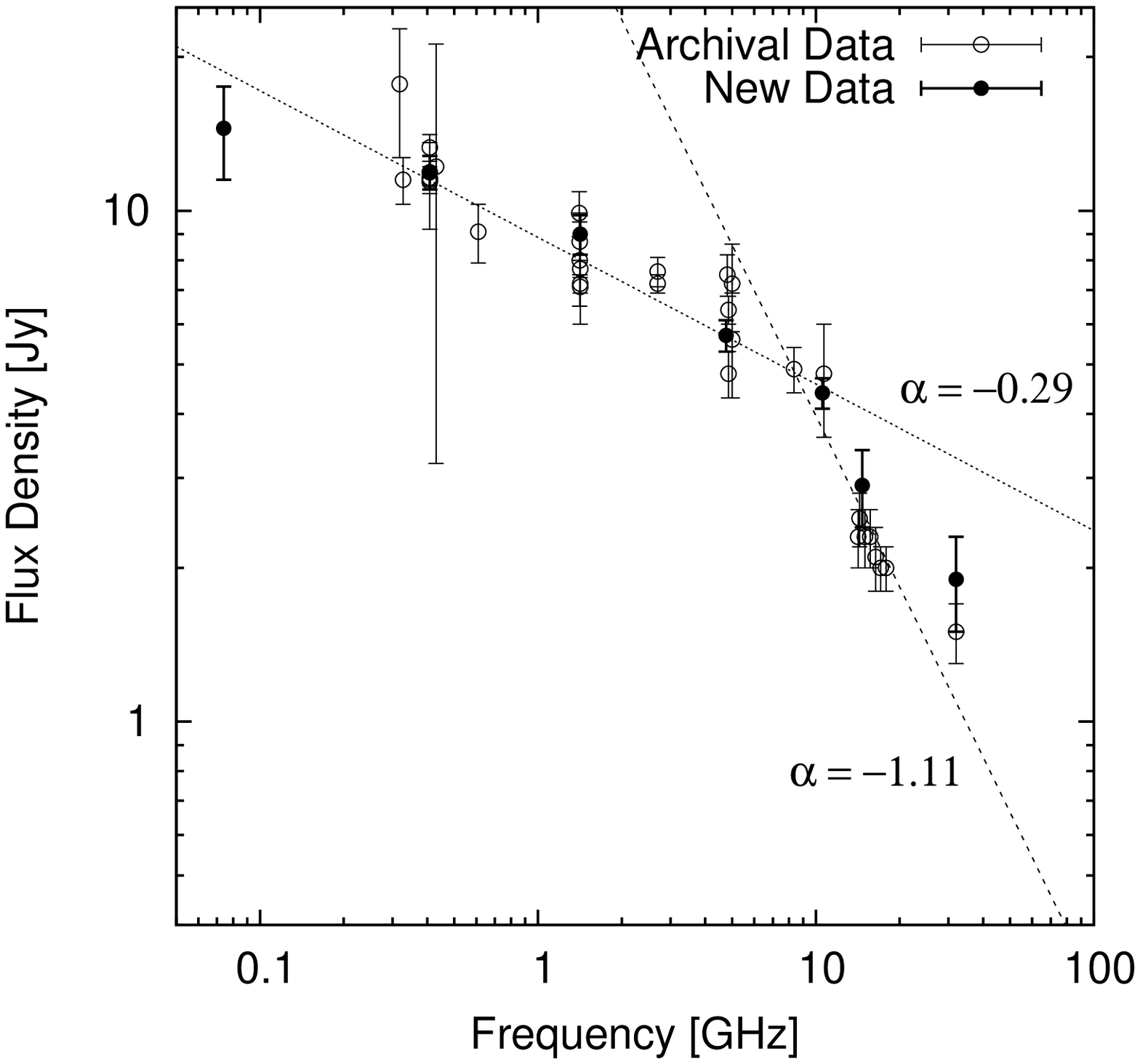}}
\caption{Integrated radio continuum spectrum of CTB\,87. Flux-density values from the
present publication are indicated by filled circles. Flux densities 
taken from the literature are shown by open circles
\citep{dick75,duin75,fant74,geld80,hurl09,koth06,lang00,mors82,mors87,pine90,reic90,reif70,sun11,wall97,weil78,wend91}.}
\label{fig:spec}
\end{figure}

Figure~\ref{fig:spec} shows the integrated radio continuum spectrum of 
CTB\,87. All flux-density values are to the best of our knowledge on the \citet{baar77} scale. The newest flux-density scale by \citet{perl17} was not used to make our results better comparable to archived studies. The ratios between the scale by \citet{baar77} and the new scale by \citet{perl17} can be found in Table~9 of \citet{perl17} and are within the calibration uncertainties anyway. The flux-density values from the present measurements are listed in
Table~\ref{tab:flux}, other flux densities were taken from the literature (see Figure caption). We fitted separate spectra to the two previously predicted high- and low-frequency parts. For the low-frequency spectrum, 
we used all flux-density measurements up to a frequency of 11~GHz, and for the high-frequency spectrum, all data above 10~GHz.
The resulting spectral indices are
$\alpha = -0.29 \pm 0.02$ for the low-frequency part and $\alpha =
-1.11 \pm 0.11$ for the high-frequency part with a break frequency of
about 8~GHz. This break frequency is somewhat lower compared to the 11-GHz break reported earlier by \citet{mors87}. This result is most likely caused by
the AMI (The Arcminute Microkelvin Imager is a radio interferometer, located at the Mullard Radio Astronomy Observatory near Cambridge, England) flux measurements in the range of 14 to 18~GHz \citep{hurl09}, which are systematically below the high-frequency measurements from single-dish telescopes. Most likely, the interferometric AMI observations suffer from missing short-spacing information. \citet{hurl09} tried to re-cover flux loss due to missing short spacings. This flux loss was estimated by matching the uv-coverage of the corresponding CGPS 1420-MHz observations. The assumption of a constant spectral index over the source may have caused the missing flux. In addition, the flux density was determined by drawing a polygon around the source, integrating the flux density inside and using the polygon edges to estimate the background. If the polygon is too small, part of the diffuse component was subtracted as background emission.

\subsection{Component Separation}

\begin{figure*}
\centerline{\includegraphics[height=0.95\textwidth,angle=-90,clip]{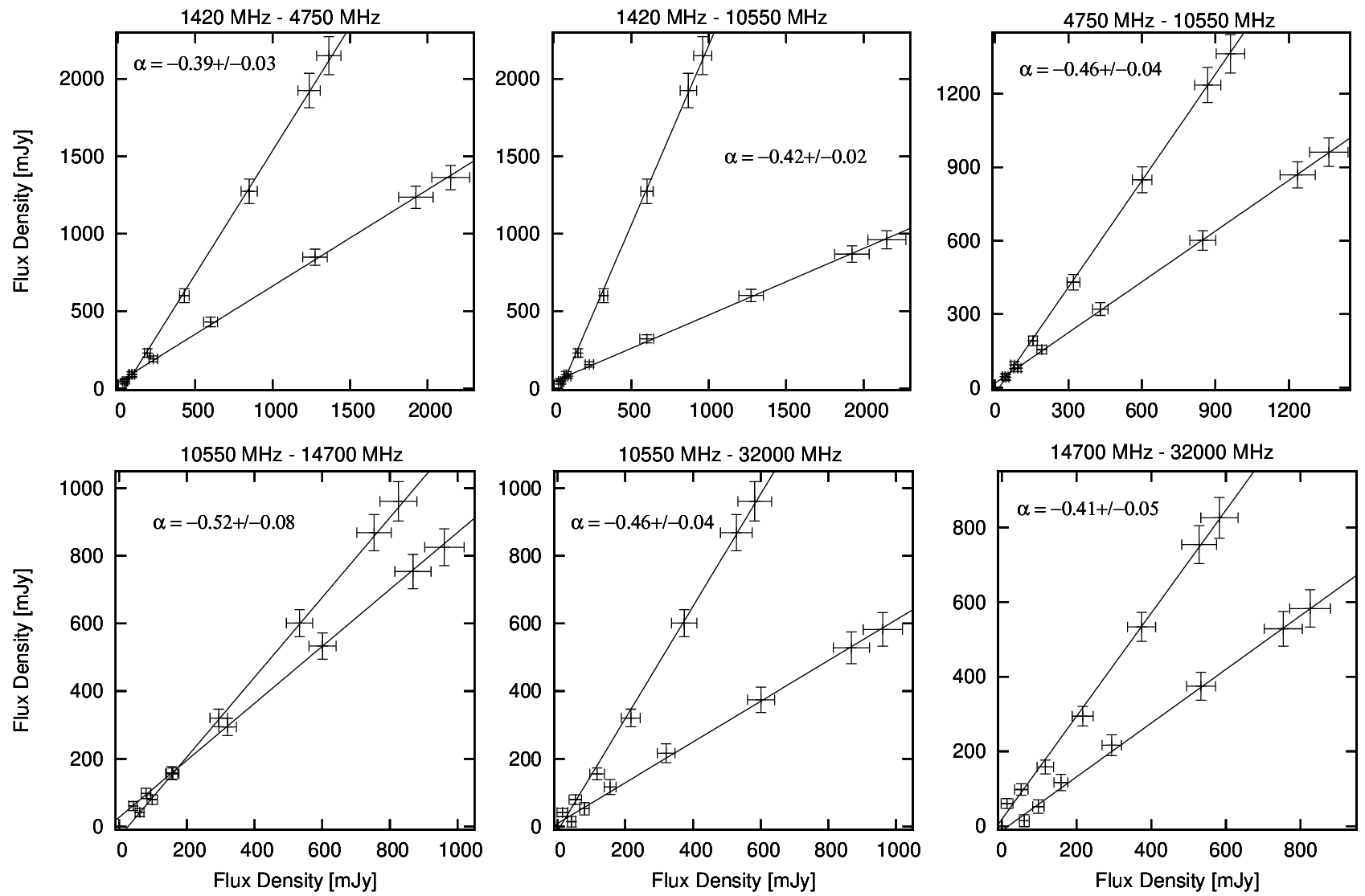}}
\caption{TT-Plots \citep{turt62} for observations of CTB\,87 at frequencies below the break frequency (upper diagrams) and above (lower diagrams). All
maps have been convolved to a common resolution of 2\farcm5. For a description of the procedure, please consult the text.}
\label{fig:tt1}
\end{figure*}

We have further investigated the spectral characteristics shown in the integrated radio continuum spectrum (Fig.~\ref{fig:spec}) by deriving spectral indices between
frequency pairs using `TT-Plots' \citep{turt62} on the radial profiles after
convolution to a common resolution of 2\farcm 5 (Fig.~\ref{fig:tt1}). The amplitudes are 
ring-averaged flux densities. For each frequency pair, we did a least-square fit of the flux densities at each frequency as a function of the flux densities at the other frequency. Both fits are displayed in each panel of Fig.~\ref{fig:tt1}. The resulting spectral indices are averaged to arrive at the final value indicated in the panels.

\begin{figure*}
\centerline{\includegraphics[width=0.93\textwidth,clip]{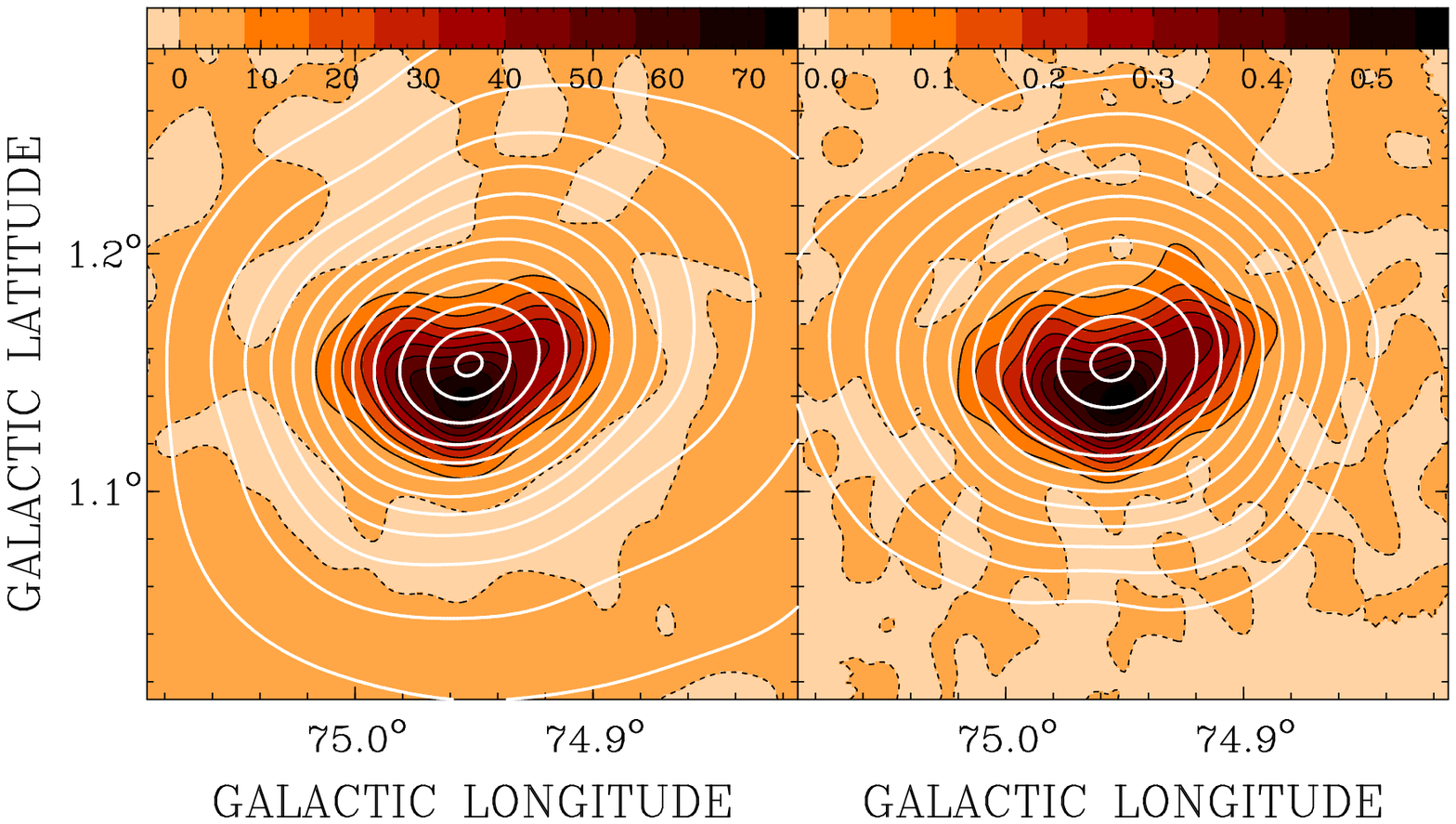}}
\caption{Total-power images of the compact component of CTB\,87 at 1420~MHz and 10.55~GHz. Map units are brightness temperature $T_{\rm b}$ in K. The compact component is shown by black contours and the diffuse component by white contours. The dashed contour indicates the zero level of the compact component. The black contour levels are indicated by the steps in the colour bar. White contours start at 1~K and run to 10~K in steps of 1~K, plus 10.8~K and 11.2~K at 1420~MHz and at 10 to 100~mK in steps of 10~mK, plus 105.5~mK at 10.55~GHz. Components were separated with the bgf method developed by \citet{sofu79} (see text).}
\label{fig:souback}
\end{figure*}

We plotted six examples in Fig.~\ref{fig:tt1}. Three
examples are shown for frequencies below the proposed break seen in Fig.~\ref{fig:spec} 
and three examples for frequencies above it. The results reveal a relatively constant
spectral index $\alpha$ between about $-$0.4 and $-$0.5 with no significant change
over the indicated break frequency. However, some of the fits have significant offsets from the origin of the diagrams.
Spectral indices based on TT-plots give
reliable results for bright objects on top of extended diffuse emission components,
which would then cause an offset from the origin of the diagram if the spectral index 
is different. Another diagnostic to detect another component with a different spectral index would be that the two fitted functions do not meet at the origin of the diagram. The latter is typically more obvious to the eye, since the offset amplitude can be relatively small for a large low surface brightness component. In Figure~\ref{fig:tt1}, this is most obvious for the TT-Plot between 10.55~GHz and 14.7~GHz, where the two fitted functions meet at about (150/150). In the TT-Plot between 14.7~GHz and 32~GHz, the two functions meet somewhere in the negatives. 

A spectral-index estimate based on flux integration includes the 
contribution of all components. If a diffuse component contains significant flux,
the two methods will give different results. This is obviously the case for CTB\,87,
which seems to consist of two different emission components. One component is rather compact, but shows high radio surface brightness. This is the well-known kidney-shaped feature. The second component is more extended and diffuse and contains much of the flux density. The diffuse component seems to be responsible for the apparent break in the integrated spectrum. It does not seem to have a well-defined outer
edge, but merges at some radius with the noise in the map. This may explain why the scatter in the integrated flux densities is so high, even though CTB\,87 is a bright radio source. Around 5~GHz,
for example, flux-density values taken from the literature vary between 4 and 8~Jy (see Fig.~\ref{fig:spec}).

\begin{figure}
\centerline{\includegraphics[width=0.45\textwidth,clip]{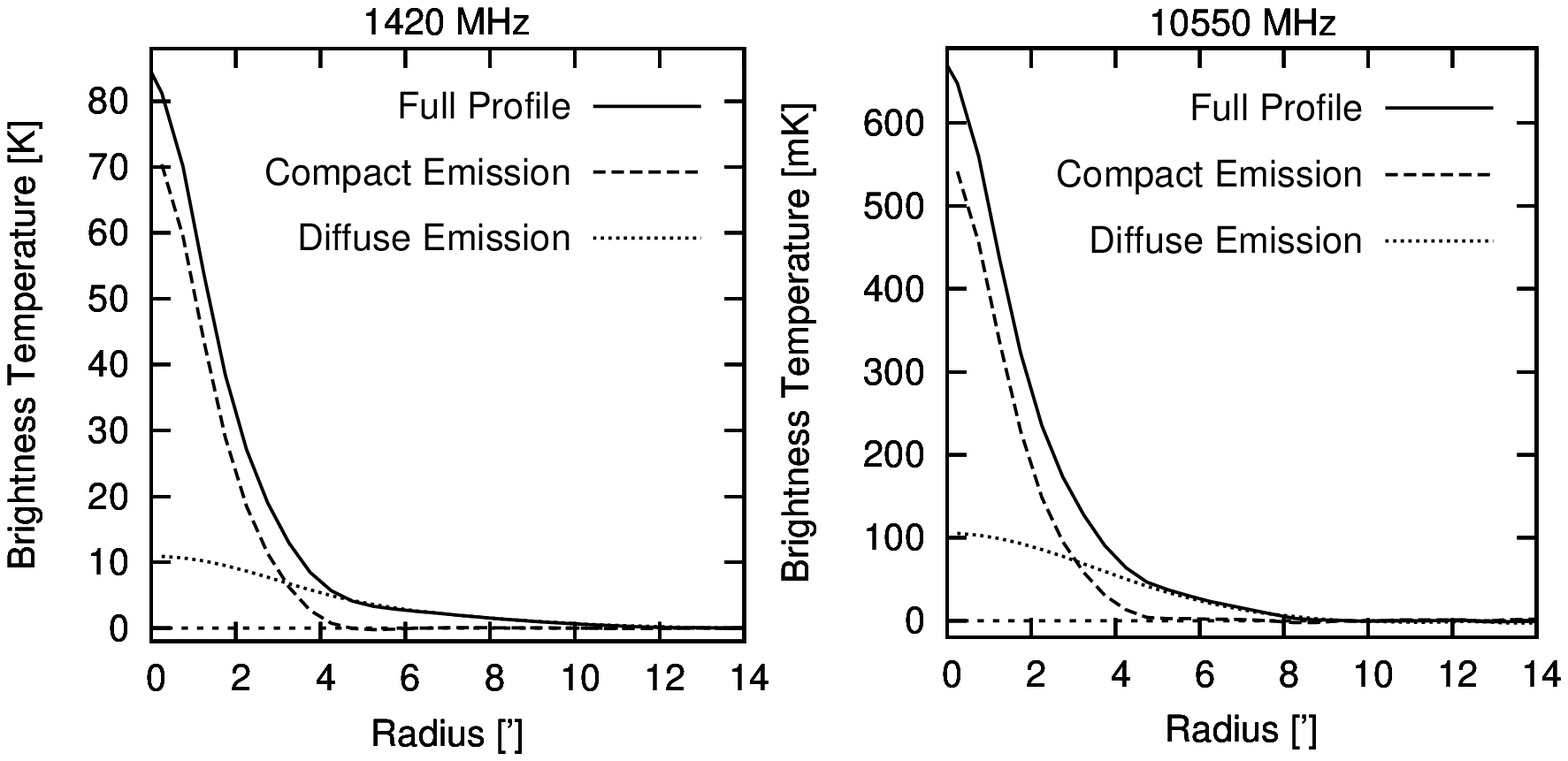}}
\centerline{\includegraphics[height=0.45\textwidth,angle=-90,clip]{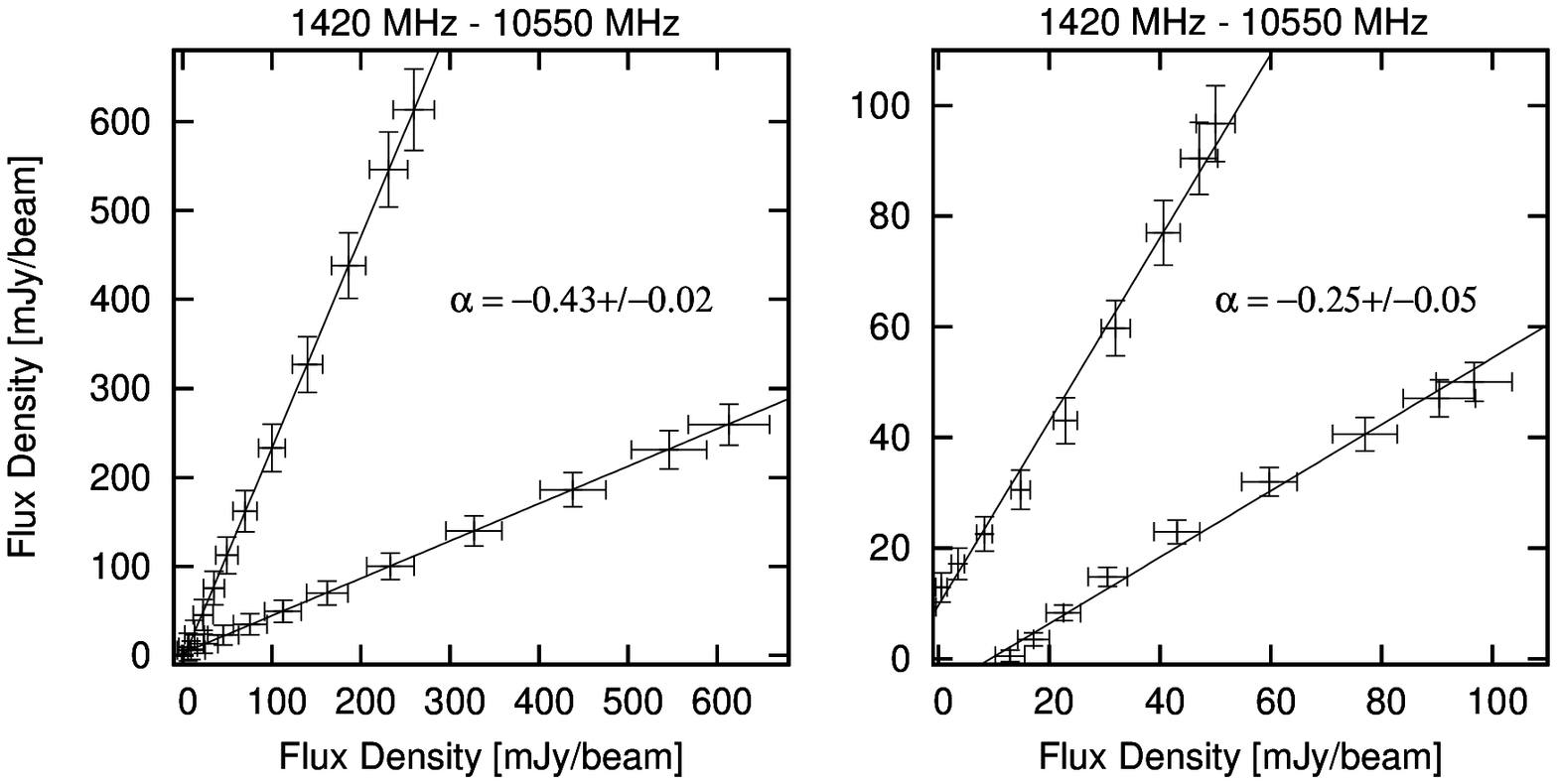}}
\caption{Top: Radial profiles of CTB\,87 at 1420~MHz (left) and 10.55~GHz (right). The
profiles of the full source, the compact source, and the diffuse emission are 
compared.
Bottom: TT-plots \citep{turt62} between radio observations of CTB\,87 at 1420~MHz and 10.55~GHz for the compact component (left) and the diffuse component (right).}
\label{fig:bgf1}
\end{figure}

In order to investigate the two components of CTB\,87 individually, we applied the
`background filtering method' (bgf, method of unsharp masking) invented by \citet{sofu79} to separate the compact (or source) component from the diffuse (or extended) component. The 408-MHz and 4.75-GHz observations have been excluded due to their low angular resolution. We applied the 
filtering or separation process first to the 1420-MHz observation from the CGPS and the Effelsberg 10550-MHz observation, since
these have covered the largest area around CTB\,87. For this procedure, we used maps
convolved to a common resolution of 1\farcm 2 and a Gaussian
filter beam of 2\farcm 5. The results are shown in Figs.~\ref{fig:souback} and \ref{fig:bgf1}. 

The radial profiles in Fig.~\ref{fig:bgf1} show that the separation was 
successful and the two components indeed represent different spatial scales.
The TT-plots created with those radial profiles confirm the spectral index
of the compact component to be between -0.4 and -0.5 as seen in Fig.~\ref{fig:tt1}. 
The diffuse component
has a significantly flatter radio spectrum with $\alpha = -0.25\pm 0.05$,
which is typical for a pulsar wind nebula.

Images of the compact component of CTB\,87 superimposed with contours from the diffuse component at 1420~MHz and 10.55~GHz are displayed in Fig.~\ref{fig:souback}. The radio peak
of the diffuse component is about $2\arcmin$ offset north from the peak of the compact component. This could indicate a different
origin of the emission, a highly structured environment or a source evolution effect. At 10.55~GHz, the diffuse
component appears slightly different in shape compared to 1420~MHz. This is likely caused by confusion with
other diffuse unrelated background emission, which is not surprising given the proximity of the
Cygnus-X region, one of the most complex features in the radio sky. This will make it in principle very 
difficult to properly characterise the diffuse component.

\begin{figure}
 \centerline{\includegraphics[width=0.48\textwidth,clip]{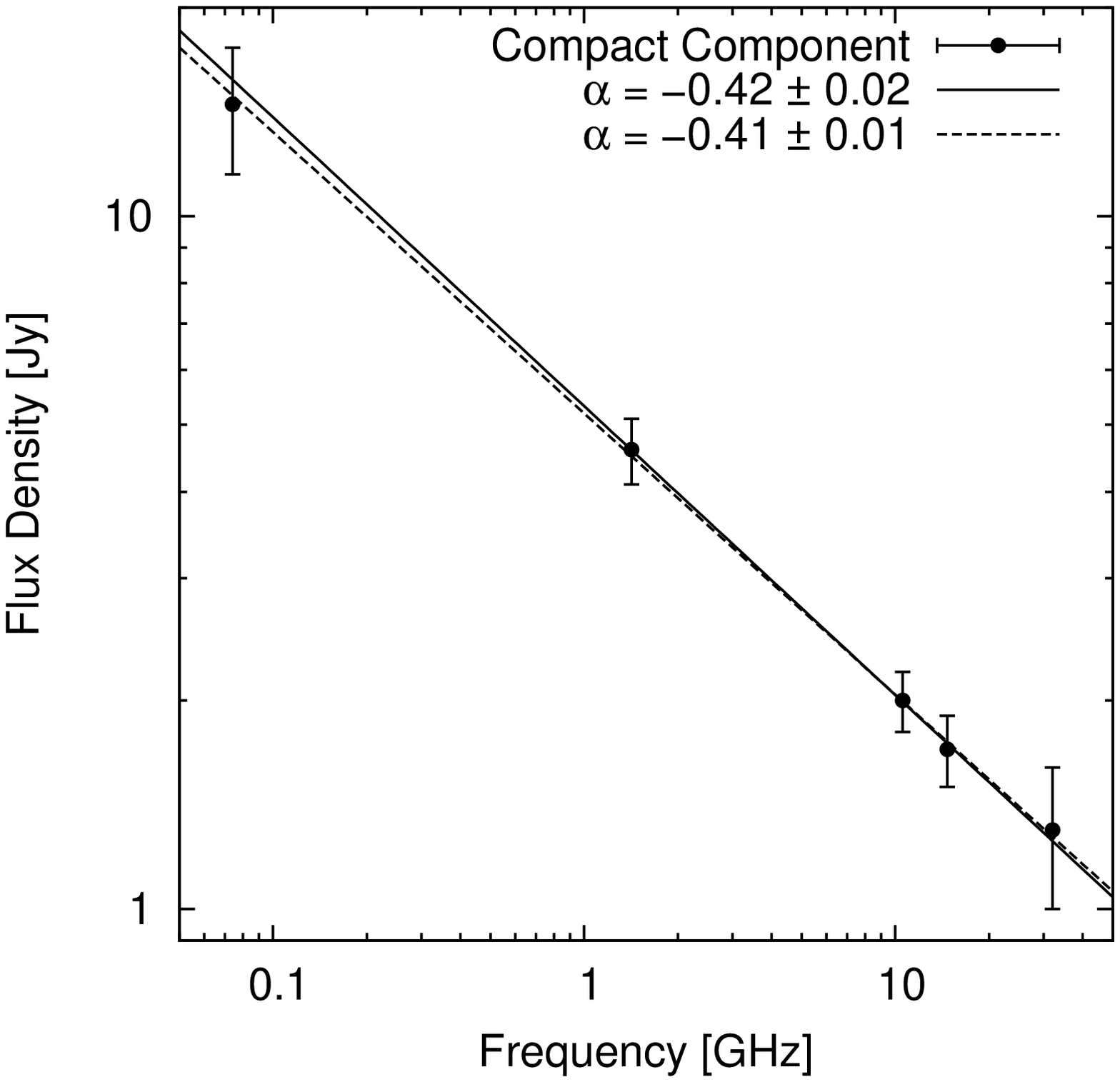}}
\caption{Integrated radio continuum spectrum of the compact component of CTB\,87. The dashed
line represents a fit to all flux values displayed and the solid line was fitted
to the flux densities at 1420~MHz, 10.55~GHz, 14.7~GHz, and 32~GHz only.}
\label{fig:speccomp}
\end{figure}

\begin{figure*}
\centerline{\includegraphics[width=0.95\textwidth,clip]{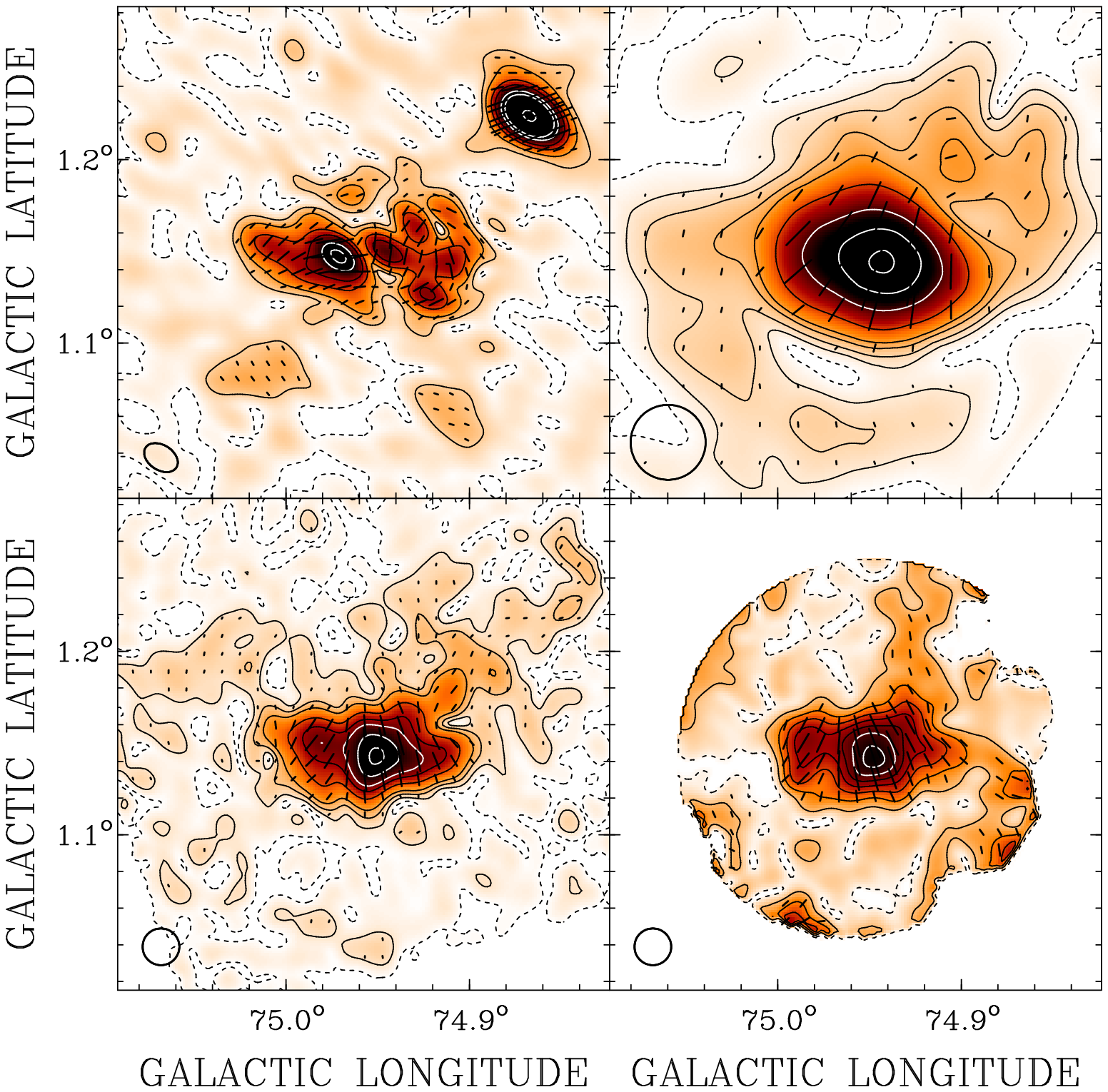}}
\caption{Images of polarised intensity of CTB\,87 at 1420~MHz (top left), 4.75~GHz (top right), 10.55~GHz (bottom left), and 32~GHz (bottom right). Vectors in observed E-field direction are overlaid. Vectors are plotted approximately half the beam size apart and their length is scaled linearly with polarised intensity. Solid contours indicate the level of polarised intensity. At high polarised intensities, white contours were used on the dark background. Contours are plotted at 0.1, 0.2, 0.4, 0.6, 1.0, 1.3, and 2.7~K at 1420~MHz, at 0.01, 0.02, 0.035, 0.05, 0.1, 0.2, 0.3, and 0.38~K at 4.75~GHz, at 40, 90, 150, 300, 450, 570, 770, and 940~mK at 10.55~GHz, and at 8 to 56~mK in steps of 8~mK at 32~GHz. The dashed contour indicates the zero level.}
\label{fig:pi}
\end{figure*}

In addition to 1420~MHz and 10.55~GHz, we also applied 
the background-filtering method with the same parameters to the observations at 
14.7~GHz and 32~GHz. Integrated flux densities of the compact component at all four 
frequencies are listed in Table~\ref{tab:flux} and displayed in 
Fig.~\ref{fig:speccomp}. In the spectrum in Fig.~\ref{fig:speccomp}, we also added 
the 74-MHz flux density from the VLSSr \citep{vlss}, assuming that the diffuse component would have been filtered out, because it results from an interferometric snap-shot survey with the JVLA. Even though the JVLA has the appropriate short spacings to cover even the diffuse component, UV coverage of the short spacings is very poor due to the low integration time. The resulting spectral index is $\alpha = -0.41 \pm 0.01$,
which confirms the results from the TT-Plots. The fitted lines in the TT-Plots intersect now with the origin of the diagram (Fig.~\ref{fig:bgf1}), indicating that a nearly perfect zero-level adjustment of the individual maps was obtained and no significant additional confusing component with a different spectral index is left.

In the TT-plot for the diffuse component between 1420~MHz and 10.55~GHz
(Figure~\ref{fig:bgf1}), there is a large offset from the origin of the diagram,
which is likely caused either by unrelated background emission at 1420~MHz
or the map at 10.55~GHz is too small.

\subsection{Linear Polarisation and Faraday Rotation}

\begin{figure*}
\centerline{\includegraphics[width=0.95\textwidth,clip]{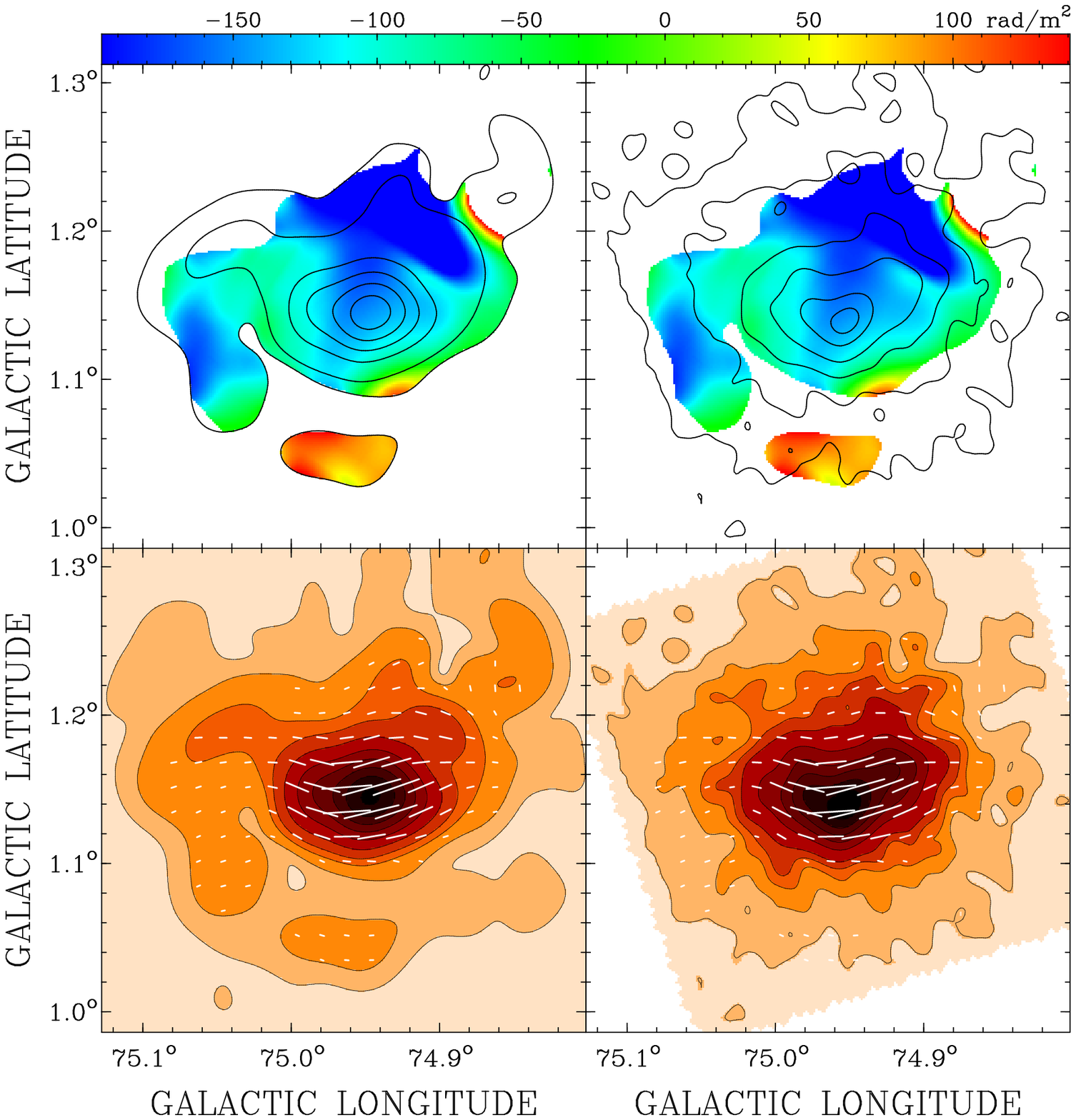}}
\caption{Top: Images of rotation measure calculated between 4.75~GHz and 10.55~GHz at a resolution of $2\farcm5$ with contours of polarised intensity (left) and total power (right) at 10.55~GHz.
Bottom: Maps of polarised intensity (left) and total power (right) at 10.55~GHz with overlaid B-vectors corrected for Faraday rotation.Vectors are plotted approximately half the beam size apart.}
\label{fig:rmmap}
\end{figure*}

Polarised-intensity images of the new radio continuum observations carried out with the Effelsberg telescope along with images from the DRAO ST are presented in Fig.~\ref{fig:pi}. The polarised intensity is calculated as described in \citet{uyan03} using the noise bias correction from \citet{ward74}. The linear-polarisation signal is strongest towards the compact kidney-shaped feature with peak percentage polarisation of up to 19\,\% (Table~\ref{tab:flux}). However, in particular at 4.75~GHz, but also at 1420~MHz and 10.55~GHz, there seems to be some faint diffuse patches of polarised emission outside the compact component related to the diffuse component as well. This emission is too faint to be properly quantified. At 32~GHz, only the kidney-shaped feature 
is covered in the map shown in Fig.~\ref{fig:pi}. 
At the higher frequencies, which are not much affected by
Faraday rotation, the E-vectors show a radial orientation relative to the
curvature of the kidney-shaped feature, indicating a magnetic
field running in tangential direction. Integrated polarised intensity and fractional
polarisation are listed in Table~\ref{tab:flux}. The fractional polarisation at
4.75~GHz and higher frequencies is very similar, while the 1420-GHz value is a
lot lower indicating significant depolarisation at that frequency. Since the angular resolution at 1420~MHz is much better than those at higher frequencies, the lower percentage polarisation cannot be caused by beam depolarisation. This
implies significant internal Faraday rotation, which then causes depolarisation at
the lower frequencies. In addition, we find some kind of `depolarisation canals' and polarisation minima at 1420~MHz, which are not seen in the higher-frequency maps. This also indicates significant internal depolarisation, caused by internal Faraday rotation.

Faraday rotation is expressed as rotation measure $RM$ and is defined through:
\begin{equation}
    \phi_{\rm obs}(\lambda) = \phi_0 + RM \lambda^2.
\end{equation}
Here, $\phi_{\rm obs}(\lambda)$ is the observed polarisation angle at wavelength $\lambda$ and $\phi_0$ is the intrinsic polarisation angle of the linearly-polarised emission. $RM$ is given in rad\,m$^{-2}$ with
\begin{equation}\label{equ:rm}
    RM = 0.81 \int \left( \frac{n_e}{\rm cm^{-3}}\right) \left( \frac{B_\parallel}{\mu{\rm G}}\right) \left( \frac{dl}{\rm pc}\right) {\rm rad\,m^{-2}}
\end{equation}
Here, $n_e$ is the electron density, $B_\parallel$ is the magnetic field parallel to the line of sight and $dl$ the path length along the line of sight.

In Fig.~\ref{fig:rmmap}, we display a rotation-measure (RM) map calculated between the observations at 4.75~GHz and 10.55~GHz, using equation 1, after we convolved the higher-frequency map to the resolution of $2\farcm5$. Since we only used areas in which the signal-to-noise is high ($8\sigma$), the error in the RMs is pretty constant at about $\pm 15$~rad\,m$^{-2}$. Any position-angle measurement has an ambiguity of any integer number of $\pi$. One cannot distinguish between a polarisation angle of $\phi$ or $\phi \pm n \cdot \pi$. Therefore, in general, the position angle must be measured at three or more frequencies in order to determine RM accurately and remove the ambiguity. However, since we use observations at high radio frequencies, the resulting RM ambiguity is very high; in our case it is about $\pm~n \cdot 1000$~rad\,m$^{-2}$. Since the integrated fractional polarisation is pretty constant from 4.75~GHz to higher frequencies, the resulting RMs of our study are far less than the ambiguity. Such high rotation measures are also not expected in our observations, neither in the foreground \citep{vane11}, nor in the SNR itself. A rotation measure of $400$~rad\,m$^{-2}$ would cause a rotation of $90\degr$ at 4.75~GHz. Therefore, RM in excess of this would cause significant line-of-sight depolarisation.

The kidney-shaped feature shows rotation measures between about $-80$~rad\,m$^{-2}$ and $-160$~rad\,m$^{-2}$ (Fig.~\ref{fig:rmmap}). The RM distribution is quite symmetric with the most negative RM close to the emission peak, getting slightly larger on either side along the curvature of the kidney. This RM distribution resembles that of the Vela PWN \citep{dods03} and the Boomerang PWN \citep{koth06b}.
\citet{koth06b} explained the RM structure by a radial or dipolar magnetic field inside the PWN, dominating the line-of-sight component. Outside the kidney, there is a partial shell in polarised intensity attached to the left edge of the kidney-shaped source moving around it to the south, which is best visible at 4.75~GHz.
The RM values are similar to the compact component in the east and getting higher up to about $+100$~rad\,m$^{-2}$ in the south.

We also used the RM map and equation 1 to correct the observed 10.55-GHz polarisation angles for Faraday rotation to obtain the intrinsic polarisation angles and magnetic field direction (Fig.~\ref{fig:rmmap}). The magnetic field
projected to the plane of the sky is tangential to the kidney-shaped feature and the southern part of the
shell-like feature. In the left part of the shell feature, it is more radial. Again,
the properties of the kidney-shaped CTB\,87 feature are very similar to the Vela PWN \citep{dods03} and the Boomerang PWN
\citep{koth06b}.

\begin{figure}
\centerline{
\includegraphics[width=0.45\textwidth,clip]{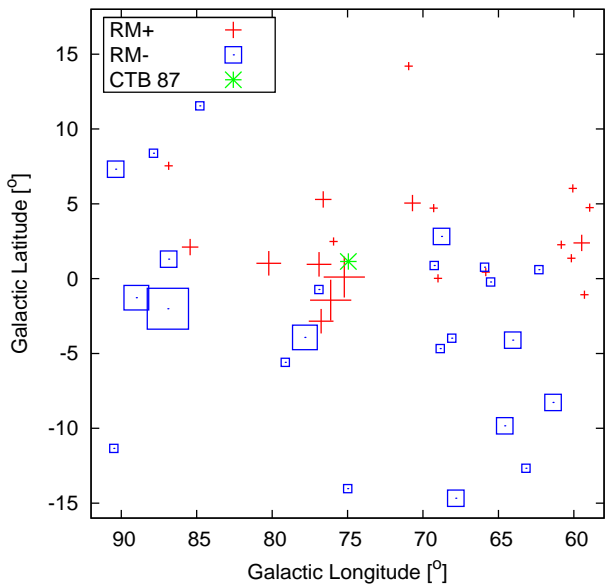}
}
\caption{Foreground RMs for pulsars within 25\degr around CTB\,87, plotted
on a grid of Galactic longitude and latitude. The data were taken from the
ATNF Pulsar Catalogue
\citep{manc05}.
Positive (RM+) and negative (RM$-$) rotation measures
are indicated, also the position of CTB\,87.
The sizes of the symbols represent the 
amplitude in RM, from 0 to $\pm 100$ (smallest symbol) to $\ge 400$~rad\,m$^{-2}$ (largest symbol) in steps of 100~rad\,m$^{-2}$.}
\label{fig:pulsar}
\end{figure}

In an attempt to separate internal from foreground Faraday rotation, we studied RMs of nearby pulsars using the ATNF (Australia Telescope National Facility) Pulsar Catalogue \citep[web page: http://www.atnf.csiro.au/research/pulsar/psrcat/]{manc05}. All - but one - pulsars within $5\degr$ of CTB\,87 have a positive rotation measure,
independent of their distance from us (see Fig.~\ref{fig:pulsar}). Their  distance range goes from 1.8~kpc to 13.2~kpc. The only exception is pulsar B\,2027+37 at ($76\fdg 90,-0\fdg 73$) in Galactic coordinates with a rotation measure of $-0.6$~rad\,m$^{-2}$ at a distance of 5.8~kpc. This is unusual, since
in this part of the Galaxy the magnetic field is supposed to point away from us, which results in negative RMs \citep[e.g.][]{brow01}. It seems likely that the Cygnus~X
region, where the pulsars with mostly positive RM are located, contains a magnetic field component pointing towards us. Another complicating factor therefore is the fact that CTB\,87 is located well beyond Cygnus~X in the Perseus spiral arm \citep{koth03}, which would then mean that along the line of sight beyond Cygnus~X there must be a field reversal.

We do not see strong small-scale fluctuations in the RM map of CTB\,87, therefore we assume that we have a relatively smooth foreground magneto-ionic medium responsible for the observed foreground Faraday rotation. To estimate the foreground RM of CTB\,87, we first determined its foreground
dispersion measure by using the recently determined electron-density model of our Galaxy by \citet{yao17} resulting in $DM \approx 200$~pc\,cm$^{-3}$. 

Given the possibility of a field reversal along the line of sight, we averaged the 4 nearby pulsars (within $5^\circ$) with a distance of $\pm 1$~kpc from CTB\,87 to get an estimate of CTB\,87's foreground magnetic field. 
This results in a mean foreground magnetic field, parallel to the line of sight, 
of $B_\parallel = +1.1 \pm 0.8~\mu$G for CTB\,87,
indicating a foreground rotation measure of about $+180 \pm 130$~rad\,m$^{-2}$.
Comparing this value with our RM map in Fig.~\ref{fig:rmmap} indicates extremely high internal
negative rotation measure for the compact part of CTB\,87. This Faraday rotation must be internal to the relic PWN. It could be caused by the mixing of the synchrotron nebula with the supernova ejecta, shock-heated by the reverse shock. 
Such a high internal Faraday rotation should completely depolarise CTB\,87 at 1420~MHz. 

The rotation measure of the nearby bright compact extra-galactic source B2013+370 points to a highly negative foreground RM.
The only available RM
for B2013+370 is listed by \citet{cleg92} as $RM = -518 \pm 6$~rad\,m$^{-2}$. From the four bands of our DRAO ST observations and the polarisation angle determined at 10.55~GHz from our Effelsberg data, we calculated $RM = -444 \pm 2$~rad\,m$^{-2}$. Therefore, we will also attempt to determine the internal Faraday rotation of CTB\,87 from its depolarisation at 1420~MHz. 

\begin{figure}
\centerline{
\includegraphics[width=0.45\textwidth,clip]{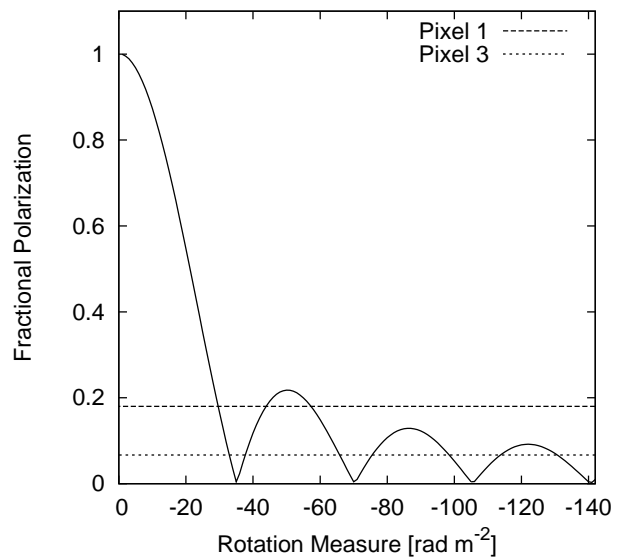}}
\caption{Fractional observed polarisation at 1420~MHz relative to the intrinsic value as a function of observed rotation measure for a Burn slab \citep{burn66}. The detected fractional polarisation at pixels 1 and 3 is indicated. Pixel 2 has no detectable polarisation at 1420~MHz. The fractional polarisation at 1420~MHz is the ratio between the percentage polarisation at 1420~MHz and the intrinsic percentage polarisation, which was calculated by averaging the percentage polarisation at 32~GHz and 10.55~GHz (see Table~\ref{tab:pixel}).}
\label{fig:depol}
\end{figure}

We used the observations with the highest resolution, which are the DRAO ST 1420-MHz observation and the Effelsberg 10.55-GHz and 32-GHz observations. We convolved the Stokes I, Q, and U maps to a common resolution of $1\farcm6 \times 1\farcm2$ with the major axis parallel to the Declination axis in the J2000 equatorial system. We chose three positions on CTB\,87 for this study, one position in the depolarisation canal (pixel 2), where PI at 1420~MHz is essentially 0, and one on PI peaks on either side of the depolarisation canal (see Fig.~\ref{fig:pi}). Pixel 1 is the highest peak in the PI map at 1420~MHz, to the left of the depolarisation canal. Pixel 3 is a PI peak to the right of it. If we assume an even mixing of synchrotron-emitting and Faraday-rotating medium in CTB\,87, we can plot depolarisation at 1420~MHz as a function of observed internal rotation measure (see Fig.~\ref{fig:depol}).

\begin{table}
    \centering
    \begin{tabular}{l|ccc}
    \hline
Pixel & 1 & 2 & 3\\
    \hline
  Pol 32~GHz & $10.7\pm 1.5$~\% & $14.2\pm 1.5$~\% & $16.4\pm 1.5$~\%  \\
  Pol 10.55~GHz & $11.4\pm 1.0$~\% & $16.2\pm 1.0$~\% & $16.5\pm 1.0$~\% \\
  Pol 1420~MHz & $2.0\pm 0.4$~\% & - & $1.1\pm 0.3$~\% \\
  RM [rad\,m$^{-2}$] & $-145\pm 34$ & $-170\pm 34$ & $-194\pm 34$ \\
  $\phi_0$ & $-15\degr$ & $+14\degr$ & $+22\degr$ \\
    \hline
  RM$_{int}$ [rad\,m$^{-2}$] & $-30$ & $-35$ & $-38$ to $-66$ \\
  RM$_{fg}$ [rad\,m$^{-2}$] & $-115$ & $-135$ & $-156$ to $-128$ \\
   \hline
  RM$_{int}$ [rad\,m$^{-2}$] & $-44$ to $-57$ & $-70$ & $-77$ to $-99$ \\
  RM$_{fg}$ [rad\,m$^{-2}$] & $-101$ to $-88$ & $-100$ & $-117$ to $-95$ \\
\hline
    \end{tabular}
    \caption{Parameters for the 3 pixels on CTB\,87 which we used for the depolarisation study. The internal and foreground rotation measures listed, RM$_{int}$ and RM$_{fg}$, represent the two possible scenarios (see text). RM was calculated between 32~GHz and 10.55~GHz.}
    \label{tab:pixel}
\end{table}

Pixel 2 is inside the depolarisation canal and represents one of the internal rotation measures at which the fractional polarisation is 0 (see Fig.~\ref{fig:depol}). Since pixel 1 has the highest fractional polarisation in the 1420-MHz map and also shows the lowest observed RM amplitude of the three pixels, it is located at a lower internal RM in Fig.~\ref{fig:depol} and pixel 3 with the highest observed RM consequently at a higher internal RM.

The fractional polarisation at 1420~MHz observed for pixel 1 constrains pixel 2 to two possible locations in Fig.~\ref{fig:depol}, the minima at $-35$ and $-70$~rad\,m$^{-2}$, since the fractional polarisation at higher internal RM is always well below the value of pixel 1. The resulting internal and foreground rotation measures have been listed in Table~\ref{tab:pixel}. While the internal RMs, due to the modelling, are quite well constrained, the foreground RMs carry the large uncertainty in the RM calculation between the observations at 32~GHz and 10.55~GHz. 

While in case 1 (pixel 2: $RM_{int} = -35$~rad\,m$^{-2}$) the internal RM is quite constant with a small gradient and the foreground shows a steep RM gradient, in case 2 (pixel 2: $RM_{int} = -70$~rad\,m$^{-2}$) it is the opposite, the internal RM is showing an RM gradient with increasing amplitude from pixel 1 to 3 and the foreground RM is relatively constant. In both cases,
the resulting average foreground RM is between about $-100$~rad\,m$^{-2}$ and $-150$~rad\,m$^{-2}$. Even though it carries a large error, this implies that the pulsars, used for the foreground RM calculation at the beginning of this subsection, see a different magneto-ionic foreground than CTB\,87 and probably also B2013+370. A highly variable foreground in the direction of Cygnus~X cannot really explain this large discrepancy. Therefore, we propose that the HI bubble in which CTB87 is located has a shell that contains a strong uniform magnetic field pointing away from us, imposing a strong negative rotation measure on the polarised emission from CTB87. 

We believe that case 2 is more realistic, since we expect a relatively constant foreground RM over the small angular scales of the PWN. Such a significant RM gradient is unexpected over such a small angular extent in the foreground Faraday rotation. However, the stellar wind bubble shell could theoretically explain the foreground gradient, if the angle between the foreground line of sight and
the foreground stellar wind bubble shell significantly changes over the face of the PWN. The gradient in the internal RM is easily explained by effects along the line of sight through the mixing of the synchrotron-emitting and Faraday-rotating media, such as a change in path length through the source or a gradual change of the magnetic field angle. 

Since smaller structures could cause changes on smaller scales, we believe that a steep RM gradient over the face of CTB\,87 is easier explained by internal Faraday rotation inside an object $14\times 8.5$~pc in size than foreground rotation caused by a shell 100~pc in diameter. Therefore, we will proceed with case 2 as the most likely scenario.

To properly describe the Faraday-depth structure along the line of sight and separate the internal and foreground RMs, we propose to make wide-band observations of CTB\,87 in L-band and S-band and use rotation measure synthesis \citep{bren05} for the foreground deconvolution. Wide-band observations at higher frequencies are not necessary, since our observations indicate the polarisation to be Faraday thin at those frequency ranges.

\section{Discussion}

\subsection{The Nature of the two Emission Components}

\subsubsection{The Compact Component}

The size of the compact component is about $7\farcm8 \times 4\farcm8$, which 
translates to spatial dimensions of 14~pc $\times$ 8.5~pc, assuming a
distance of 6.1~kpc \citep{koth03}. This compact, kidney-shaped component was identified by \citet{math13} and later confirmed by \citet{gues19} as a relic PWN, left behind by the pulsar after its wind interacted with the reverse shock of the supernova blast wave. This reverse shock was generated when the forward shock ran into the molecular cloud complex described by \citet{koth03} and later confirmed by \citet{liu18}. Another strong indicator for a relic-PWN interpretation is the presence of high internal Faraday rotation (see section 3.3). When the reverse shock interacts with the pulsar wind, it mixes the synchrotron-emitting nebula with the shock-heated ejecta. Only the thermal electrons in the ejecta can produce significant internal Faraday rotation. \citet{gues19} found an upper limit of 0.05~cm$^{-3}$ for the electron density in their \textit{XMM-Newton} observations based on a lack of any thermal X-ray emission. If we use this electron density, the width of the radio relic of 8.5~pc as a lower limit of the path length through the Faraday-rotating medium and its length of 14~pc as an upper limit, we can calculate the magnetic field necessary to produce the observed rotation measure. In section 3.3, we determined a maximum for the internal Faraday rotation of about $-100$~rad\,m$^{-2}$ for the second scenario (see Table~\ref{tab:pixel}). 

Rotation measure observed from a nebula in which the synchrotron-emitting medium is intermixed with the Faraday-rotating medium is typically about half the RM that would be imposed on linearly polarised background emission \citep{burn66}. Assuming that we are looking along the magnetic field at the maximum of $-100$~rad\,m$^{-2}$ together with equation~\ref{equ:rm} results in internal magnetic fields between $-350$ and $-580~\mu$G, using the width as a lower and the length as an upper limit for the path length. This is a very high magnetic field, but in magnitude almost as high as in other evolved PWNe that must have interacted with the reverse shock already, such as DA\,495 with 1.3~mG \citep{koth08} and the Boomerang PWN in the SNR G106.3+2.1 with 2.3~mG \citep{koth06b}. This indicates that the reverse shock has a significant impact on the magnetic field inside the relic pulsar wind nebula. For the putative pulsar that was created in the original supernova explosion, \citet{math13} determined an ambient magnetic field of $55\mu{\rm G}$ from the location of the termination shock,
so that the pulsar must have left the kidney-shaped radio nebula a long time ago.

The steep radio continuum spectrum of the compact PWN is very similar in its properties to other evolved PWNe like DA\,495 \citep[$\alpha = -0.45$]{koth08}, G76.9+1.0 \citep[$\alpha = -0.62$]{land93,land97} and the 
PWN G141.2+5.0 \citep[$\alpha = -0.7$]{koth14}. All of these show unusually steep radio continuum spectra when compared to other PWNe, which typically have spectral indices between 0.0 and $-$0.3. We believe this steep spectrum can be explained by the shock acceleration of relativistic particles by the reverse shock, making this nebula brighter in the radio regime and producing a radio continuum spectrum more similar to a shell-type SNR. This
would also have significantly compressed and enhanced the magnetic field we found inside the compact PWN. This magnetic field is found to be parallel to the reverse shock front projected to the plane of the sky.

We propose that the steep radio continuum spectrum, together with the strong magnetic field and the high internal Faraday rotation are all typical characteristics of highly evolved PWNe and are the result of strong interaction of the pulsar wind with the supernova reverse shock.

\subsubsection{The Diffuse Component}

The diffuse component seems to be almost perfectly circular at 10.55~GHz, but 
getting more elliptical at 1420~MHz (see 
Figure~\ref{fig:souback}). This is likely caused by confusion with smooth
unrelated extended emission at 1420~MHz. The diffuse component also does not
have a well-defined outer edge. It seems to disappear into the
noise at some radius.

We interpret the diffuse component of CTB\,87 as the SNR's undisturbed PWN 
inside the expanding forward shock of the supernova, most likely inside the ISM cavity or stellar
wind bubble found by \citet{liu18}. Its shape and location, 
the linearly-polarised emission, and the fact that it has no sharp outer boundary, leave no other possible explanation. Its diffuse nature with no outer edge indicates that
the reverse shock of the supernova did not interact with it yet, which means that the supernova
shock wave in this part of the SNR is still expanding freely, which may explain why no radio emission from this shell has been found yet.

The diffuse component is about {17\arcmin} in diameter (see Fig.~\ref{fig:bgf1}, top right panel), which translates to a spatial extent of 30~pc at a distance of 6.1~kpc. With 
$\alpha = -0.25 \pm 0.05$ (Fig.~\ref{fig:bgf1}), it has a much flatter
radio continuum spectrum than the compact component. This flat spectrum is 
more typical for a PWN. Since the compact source 
shows a straight radio spectrum between at least 74~MHz and 32~GHz, the
observed break in the combined spectrum must be somehow related to the
diffuse component. From our observations, we cannot decide whether the break
is real or not. However, given the uncertainty in its size and the large spread
of flux densities in the combined synchrotron spectrum that can be found in the
literature, we propose that this break is not real. The problems can be
illustrated by looking at the flux-density values determined from the CGPS 1420-MHz
data in the CGPS SNR catalogue \citep{koth06} and this study. For the SNR
catalogue, a polygon, based on its assumed diameter, was used to estimate the background
emission. Here, we used the radial profile and integrated all flux up to the first 
minimum. This results in a flux density difference of almost 2~Jy between those two studies
using the same data set.

\begin{figure}
\centerline{\includegraphics[width=0.49\textwidth,clip]{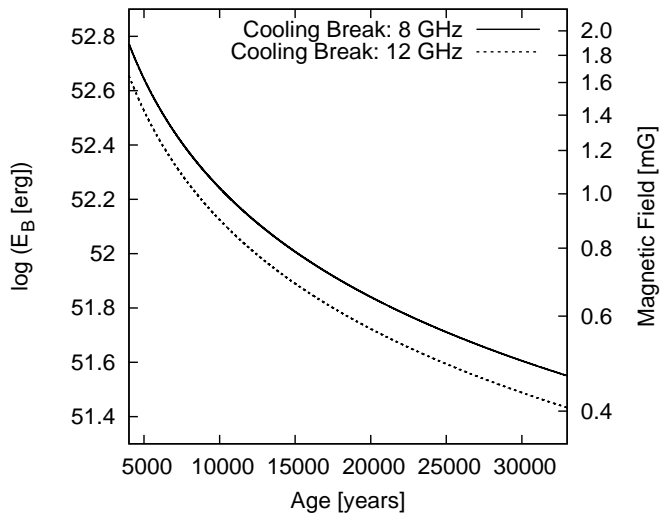}}
\caption{Required energy and magnetic field strength for the diffuse emission component as a function of time since the supernova explosion for two different cooling break frequencies.}
\label{fig:bener}
\end{figure}

Since the compact component does not show any evidence of a break in the frequency range displayed in Fig.~\ref{fig:speccomp}, any possible break in the radio synchrotron spectrum cannot be related to the intrinsic break of the relativistic electron spectrum, injected by the pulsar. If it exists, it must be the synchrotron cooling break of the diffuse component. The cooling break frequency, $\nu_c$, of the synchrotron emission spectrum is defined by \citep{chev00}:
\begin{equation}\label{equ:cool}
   \nu_c [{\rm GHz}] = 1.187\cdot B_{\rm PWN}^{-3} [{\rm G}] \cdot t^{-2} [{\rm yr}],
\end{equation}
Using equation~\ref{equ:cool}, we can estimate the magnetic field inside the diffuse component as a function of time for a cooling break frequency of $v_c = 8(12)$~GHz. This can be turned into the required energy in the magnetic field $E_{\rm B}$ as a function of time \citep[e.g.][equation 2]{koth06b}. A quick look at Fig.~\ref{fig:bener} shows that the energy requirements for a cooling break at such a low frequency are quite unreasonable. Therefore the apparent break in the radio continuum spectrum at 8(12)~GHz cannot be physical. Highly-sensitive radio continuum observations at high radio frequencies are required to properly characterise the diffuse component. However, given the high confusion in the Cygnus-X area, this might prove very challenging.

\subsection{Nature and Evolution of the CTB\,87 system}

As shown before, CTB\,87 consists of two different emission components at radio frequencies. The compact kidney-shaped component was identified as a relic PWN and the diffuse extended component is the undisturbed part of the PWN expanding inside a stellar wind bubble. This is of course not to be taken literally. There is no sharp border between those components. At the edges of the relic there is a gradual transition with the mixing of the synchrotron emitting components. But the edges of the molecular clouds are quite sharp \citep{liu18}, which makes this transition region rather small, so that we were able to separate those components spatially.

The energy-loss rate of a young pulsar, which is the main energy source for its PWN, drops quite dramatically during its early evolution. In addition, the expansion of the PWN is dominated by its environment, which typically are the expanding ejecta, and not the movement of the pulsar. This means the PWN is not `following' the pulsar. It expands from the location it was ejected. Therefore the radio peak of the diffuse emission component should be at or very close to the location of the original supernova explosion. With that in mind, we can put limits on the PWN's age, the mass of the ejecta, and the explosion energy by following the evolution of a PWN inside an expanding shell-type supernova remnant as outlined by \citet{koth17}. The equations are based on the expansion law of a PWN inside the supernova ejecta by \citet{chev04} and the equations that describe the expansion of the blast wave and the kinematics of the reverse shock by \citet{mcke95}. The expansion law of a PWN described by \citet{chev04} assumes a constant energy-loss rate for the pulsar. We adapted this equation to include the change of the energy-loss rate assuming a dipolar magnetic field \citep[e.g.][]{paci73}. This results in:
\begin{equation}\label{equ:pwn}
  R_{\rm PWN} ({\rm pc}) = 3.228\times 10^{-4} \left(\frac{\dot{E}_0 * \tau_0}{\tau_0 + t}\right)^{0.254} E_0^{0.246} M_0^{-0.5} t^{1.254}
\end{equation}
Here, $\dot{E}_0$ is the intrinsic energy-loss rate of the pulsar in $10^{38}$~erg\,s$^{-1}$ and $\tau_0$ its intrinsic characteristic age in years. $E_0$ is the explosion energy of the supernova in $10^{51}$~erg, $M_0$ is the ejecta mass in M$_\odot$, and $t$ is the age of the system in years.

To calculate the intrinsic values $\dot{E}_0$ and $\tau_0$, we need an estimate of their current values. Since we did not detect any pulsation from the neutron star, we use the X-ray luminosity of the neutron star and the PWN \citep{gues19} together with the empirical results of \citet{li08} to get estimates for the current values of $\dot{E}$ and $\tau$. \citet{gues19} estimated $\dot{E} = 1.8 \times 10^{36}$~erg\,s$^{-1}$ and $\tau = 34.5$~kyr from the pulsar luminosity and $\dot{E} = 7.5 \times 10^{36}$~erg\,s$^{-1}$ and $\tau = 11.5$~kyr from the PWN luminosity, respectively. For our study, we adopted the average values of $\dot{E} = 3.6 \times 10^{36}$~erg\,s$^{-1}$ and $\tau = 20$~kyr. 

We can also get an estimate of the current energy-loss rate of an unseen pulsar based on the current radio surface brightness of its pulsar wind nebula. \citet{koth98} found a strong correlation between a PWN's radio surface brightness at 1~GHz, its diameter, and the current energy-loss rate of the central pulsar. From our study, we estimated this radio surface brightness of the two radio components of CTB\,87 to be $2.1 \times 10^{-20}$~Watt\,m$^{-2}$\,Hz$^{-1}$\,sr$^{-1}$ at a diameter of 11~pc and $2.5 \times 10^{-21}$~Watt\,m$^{-2}$\,Hz$^{-1}$\,sr$^{-1}$ at a diameter of 30~pc for the relic and undisturbed PWNe, respectively. This allows an estimate for the current energy-loss rate of $8.5 \times 10^{36}$~erg\,s$^{-1}$ and $2.8 \times 10^{36}$~erg\,s$^{-1}$. The relic PWN is the result of the interaction of the pulsar wind with the reverse shock of the supernova explosion, therefore the $\dot{E}$ extrapolated from its radio surface brightness can only be considered as an upper limit. The $\Sigma-D-\dot{E}$ correlation clearly assumes no energy loss inside the PWN, as evidenced by the exponents of 1 and $-1$ for $\dot{E}$ and $D$ in the correlation function of \citet{koth98}. Therefore, the $\dot{E}$ extrapolated from the undisturbed PWN must be considered as a lower limit to the real value. Obviously, the estimates of the current energy-loss rates of the unseen pulsar by radio and X-ray measurements agree very well within uncertainties.

\begin{figure}
\centerline{\includegraphics[width=0.48\textwidth,clip]{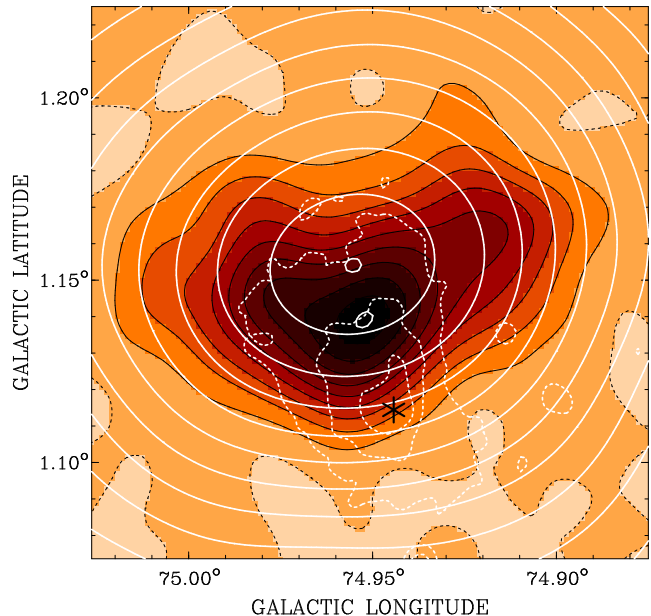}}
\caption{Total-power image of the compact component of CTB\,87 at 10.55~GHz. The diffuse radio component is indicated by the solid white contours. Contour levels are the same as in Fig.~4. The location of the putative neutron star found by \citet{math13} is marked by the black asterisk. Dashed white contours indicate the integrated X-ray emission from the latest \textit{XMM} observations \citep{gues19}.}
\label{fig:path}
\end{figure}

In Fig.~\ref{fig:path}, we 
overlay contours of the diffuse PWN on an image of the relic PWN. On this figure, we marked  the two radio peaks and the location of the putative neutron star detected with \textit{CHANDRA} and \textit{XMM-Newton} \citep{math13,gues19}. We propose that the two radio peaks and the current location of the pulsar, created in the supernova explosion,
mark the trajectory of this pulsar since its birth. 
The radio peak of the diffuse emission marks the location of the supernova explosion. While the radio peak of the compact component should indicate the place with the strongest interaction between the pulsar wind and the reverse shock. This should have happened when the reverse shock made its closest approach to the pulsar. This means that the pulsar was hit by the reverse shock at about a third of its current age.

We will now follow the evolution of the CTB\,87 system to put constraints on the initial parameters of the supernova explosion and the pulsar, and get an idea about the three-dimensional configuration of the environment. In the end, we will attempt to identify any candidates for a radio shell that marks the current location of the forward shock.

\subsubsection{The undisturbed PWN}

\begin{figure}
\centerline{\includegraphics[width=0.49\textwidth,clip]{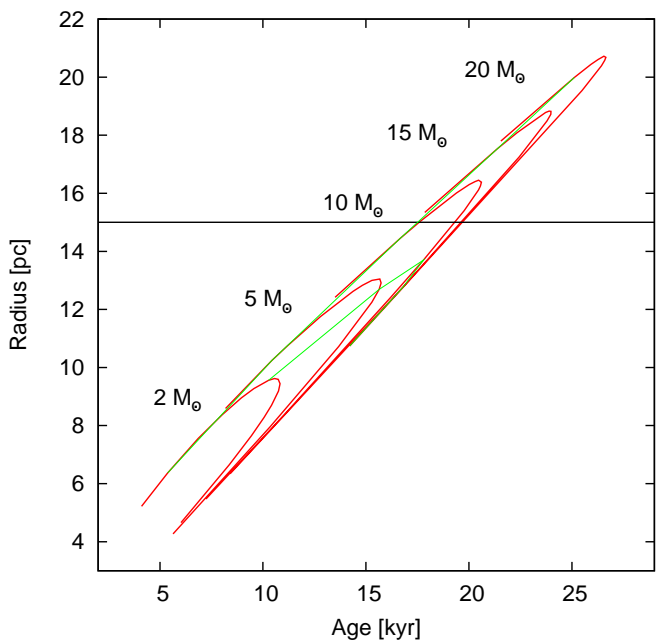}}
\caption{The radius at which the reverse shock collides with the expanding PWN as a function of time for a supernova expanding inside a stellar wind bubble of number density $0.02$~cm$^{-3}$. We plotted this function for different ejecta masses for explosion energies above $10^{49}$~erg. The green lines mark constant explosion energies of $10^{50}$~erg and $10^{51}$~erg from right to left. The radius of the diffuse component of 15~pc is indicated.}
\label{fig:diff-rad}
\end{figure}

Assuming that the diffuse PWN is indeed expanding freely inside the cavity or stellar wind bubble found by \citet{liu18}, we can calculate the time and radius of the PWN, when the reverse shock starts to interact with its edge. Inside this cavity, we adopt a number density of 0.02~atoms\,cm$^{-3}$, typical for the interior of a stellar wind bubble \citep{weav77}. To describe the hydrodynamics, we use the equations that describe the expansion of the blast wave and the kinematics of the reverse shock by \citet{mcke95}. For each combination of ejecta mass $M_0$ and explosion energy $E_0$, we can extrapolate the intrinsic values for the pulsar's energy-loss rate $\dot{E}_0$ and characteristic age $\tau_0$ using equation~\ref{equ:pwn} with the current values for $\dot{E}$ and $\tau$, the current radius of 15~pc for the PWN, and assuming that the pulsar has a pure dipolar magnetic field. 
In Fig.~\ref{fig:diff-rad}, we display the radius of the PWN at the time, when the reverse shock collides with its edge as a function of age for different ejecta masses and explosion energies.

For each ejecta mass, the curve for the collision radius of the PWN with the reverse shock describes a loop with a central maximum radius. To the left of that maximum are the higher-energy explosions, where the reverse shock started to move back inwards, before it collides with the PWN. To the right, the PWN actually catches up with the reverse shock that is still moving outwards. We can immediately extract two sharp lower limits for age and ejecta mass from these calculations. The ejecta mass of the supernova must have been at least 8~M$_\odot$ and the age of the system must be higher than 17,000~yr. Otherwise the PWN could not have reached a radius of 15~pc without colliding with the reverse shock of the supernova. The minimum explosion energy is $1.5 \times 10^{50}$~erg, otherwise the expanding PWN would have caught up with the reverse shock at a radius smaller than 15~pc. Those characteristics already favour a type II supernova explosion, consistent with the X-ray detection of the neutron star candidate powering CTB\,87 \citep{math13,gues19}.

\subsubsection{The radio relic PWN}

We can now use the location, where the reverse shock made its closest approach to the neutron star and the distance of the molecular cloud from the place of the supernova explosion together with the modelling of the undisturbed PWN in the previous subsection to put further constraints on the initial explosion parameters and the current location of the blast wave. The molecular cloud complex with which CTB\,87 is likely interacting is located to the south and southeast in Fig~\ref{fig:path}. It must be located outside the relic PWN and the diffuse X-ray emission in that direction. Therefore, the minimum distance from the location of the SN is 6~pc projected to the plane of the sky. 

We simulated the expansion of the blast wave in the direction of the molecular cloud complex assuming a minimum distance of 6~pc and using the equations by \citet{mcke95}. According to the pulsar trajectory in Fig.~\ref{fig:path}, the reverse shock has to make its closest approach to the pulsar at about one third of its current age $t$. We also cross-correlated with the simulations in Fig.~\ref{fig:diff-rad} to make sure that the PWN has a diameter of 15~pc at the current age and has not interacted with the reverse shock inside the bubble yet. The mean density inside the molecular cloud complex was taken to be 80~cm$^{-3}$ \citep{liu18}.

\begin{table}
    \centering
    \begin{tabular}{ccccccc}
    \hline
$M_0$ & $E_0$ & $\tau_0$ & $\dot{E}_0$ & $t$ & $R_{cav}$ & $R_{CO}$\\
$M_\odot$ & $10^{50}$erg & yr & $10^{38}$erg\,s$^{-1}$ & yr & pc & pc \\
    \hline
10 & 5.5 & 1912 & 3.9 & 18088 & 29.0 & 11.6 \\
12 & 7.0 & 1759 & 4.7 & 18241 & 30.5 & 11.9 \\
15 & 9.0 & 1535 & 6.1 & 18465 & 32.1 & 12.2 \\
20 & 13 & 1318 & 8.3 & 18682 & 34.6 & 12.8 \\
    \hline
10 & 7.0 & 2219 & 2.9 & 17781 & 30.4 & 12.8/11.1 \\
12 & 8.5 & 1992 & 3.6 & 18008 & 31.7 & 13.1/11.3 \\
15 & 11 & 1757 & 4.7 & 18243 & 33.4 & 13.4/11.6 \\
20 & 15 & 1459 & 6.8 & 18541 & 35.6 & 13.9/12.0 \\
    \hline
    \end{tabular}
    \caption{Parameters resulting from two simulations of the evolution of the CTB\,87 system (see main text). The top results assume that the molecular cloud is at the same distance from us as the supernova explosion. The bottom results assume a shift along the line of sight at an angle of 30\degr with the plane of the sky. Displayed are the ejecta mass $M_0$, explosion energy $E_0$, intrinsic characteristic age $\tau_0$ and intrinsic energy-loss rate $\dot{E}_0$ of the pulsar, the age of the system $t$, the radius of the blast wave inside the HI cavity $R_{cav}$, and the radius of the blast wave inside the molecular cloud $R_{CO}$. In the bottom results for $R_{CO}$, the actual radius and the radius projected to the plane of the sky are displayed.}
    \label{tab:charac}
\end{table}

\begin{figure}
\centerline{\includegraphics[width=0.45\textwidth,clip]{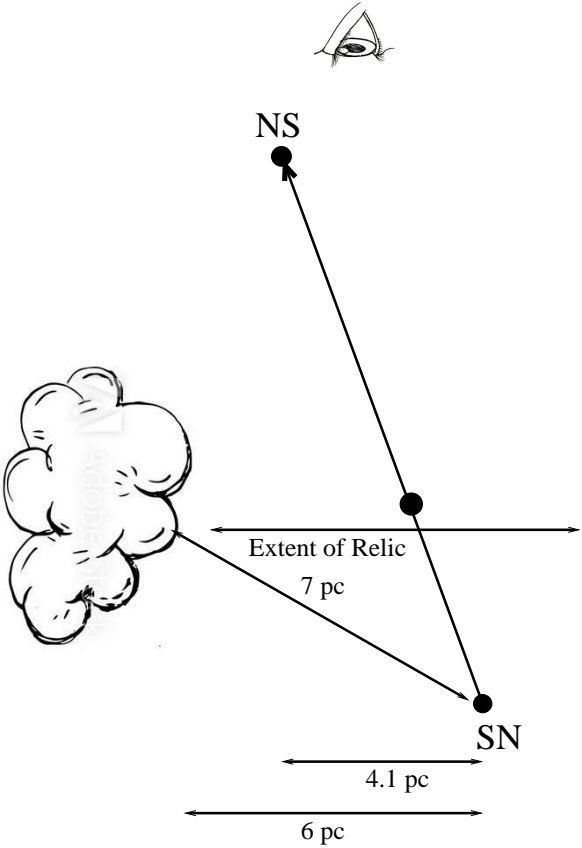}}
\caption{Sketch of the complex system of CTB\,87 outlining distances and the different components. The view is perpendicular to the plane of the sky. In this scenario, the molecular cloud is placed at a distance of 7~pc from the location of the supernova explosion. The size of the molecular cloud is not to scale. NS indicates the current location of the neutron star.}
\label{fig:sketch}
\end{figure}

The results are given in Table~\ref{tab:charac}. We explore two possible scenarios. In the first case, we assume that the centre of this molecular cloud complex, seen from the place where the supernova exploded, is at the minimum distance of 6~pc. In the second scenario, the distance is 7~pc, which is the case if the centre of the molecular cloud is shifted along the line of sight at an angle of {30\degr} with the plane of the sky (see also sketch in Fig.~\ref{fig:sketch}). The shift along the line of sight must be towards us, since the pulsar must have interacted with the reverse shock early on and is moving towards us at a relatively small angle with the line of sight \citep{gues19}. 

In summary, we find a possible range of ejecta masses between 10 and 14 M$_\odot$. The upper limit comes from the maximum electron density of $n_e = 0.05$ cm$^{-3}$. { The ejecta mass has to be less than 15~M$_\odot$ or the electron density must be higher than 0.05~cm$^{-3}$, assuming an even distribution of the ejecta between the location of the explosion and the molecular cloud \citep{gues19}. An uneven distribution would require an even lower ejecta mass.} Below 10 M$_\odot$ our simulations cannot converge. This results in an explosion energy of $E_0 = 7\pm 2 \times 10^{50}$ erg for the supernova and an age of around 18,000 yr. The intrinsic parameters of the pulsar result to $1600 \le \tau_0 \le 2200$ yr and $3 \times 10^{38} \le \dot{E}_0 \le 6 \times 10^{38}$ erg\,s$^{-1}$.

\subsection{The radio shell of CTB\,87}

\begin{figure}
\centerline{\includegraphics[width=0.45\textwidth,clip]{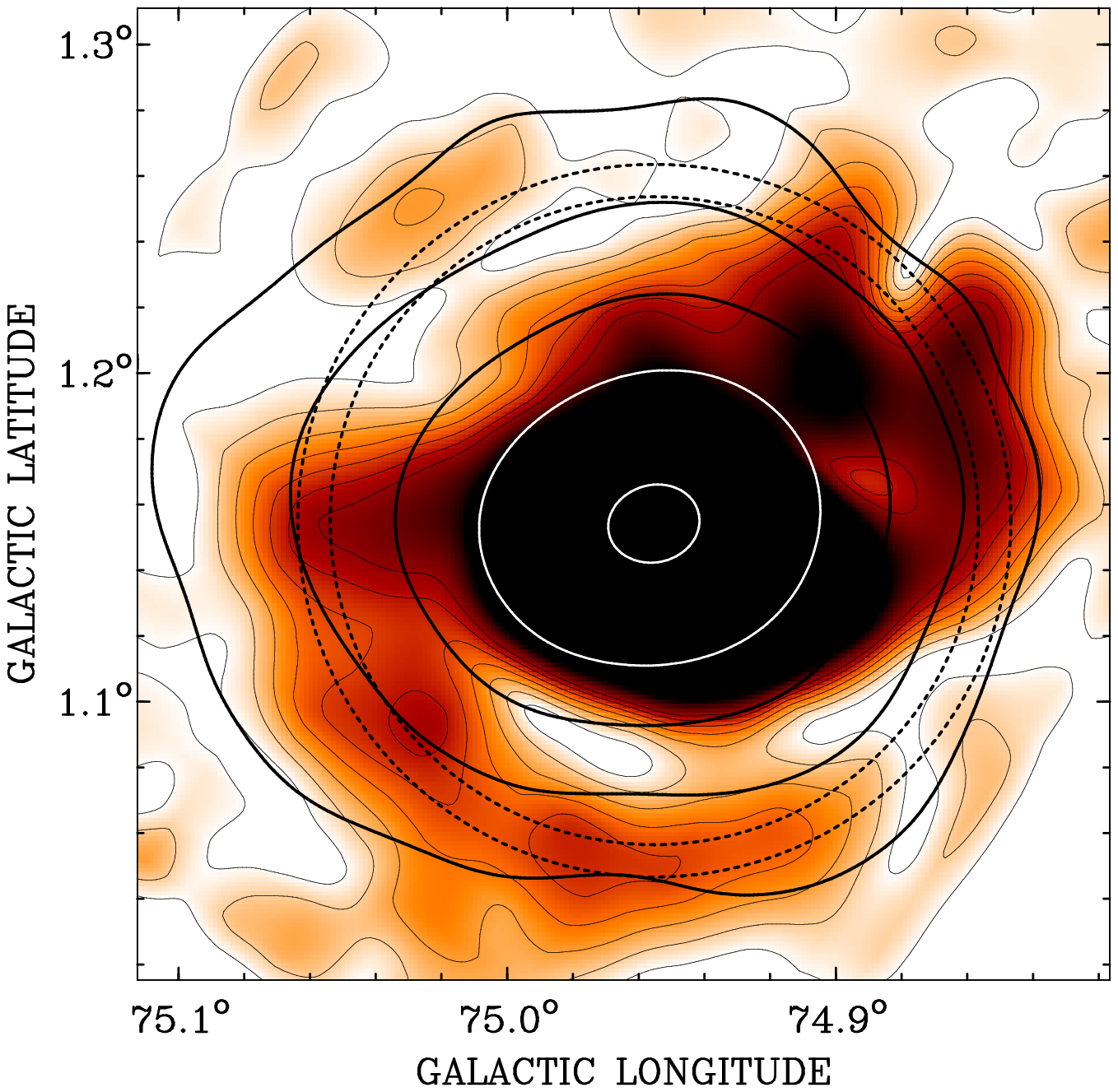}}
\centerline{\includegraphics[width=0.45\textwidth,clip]{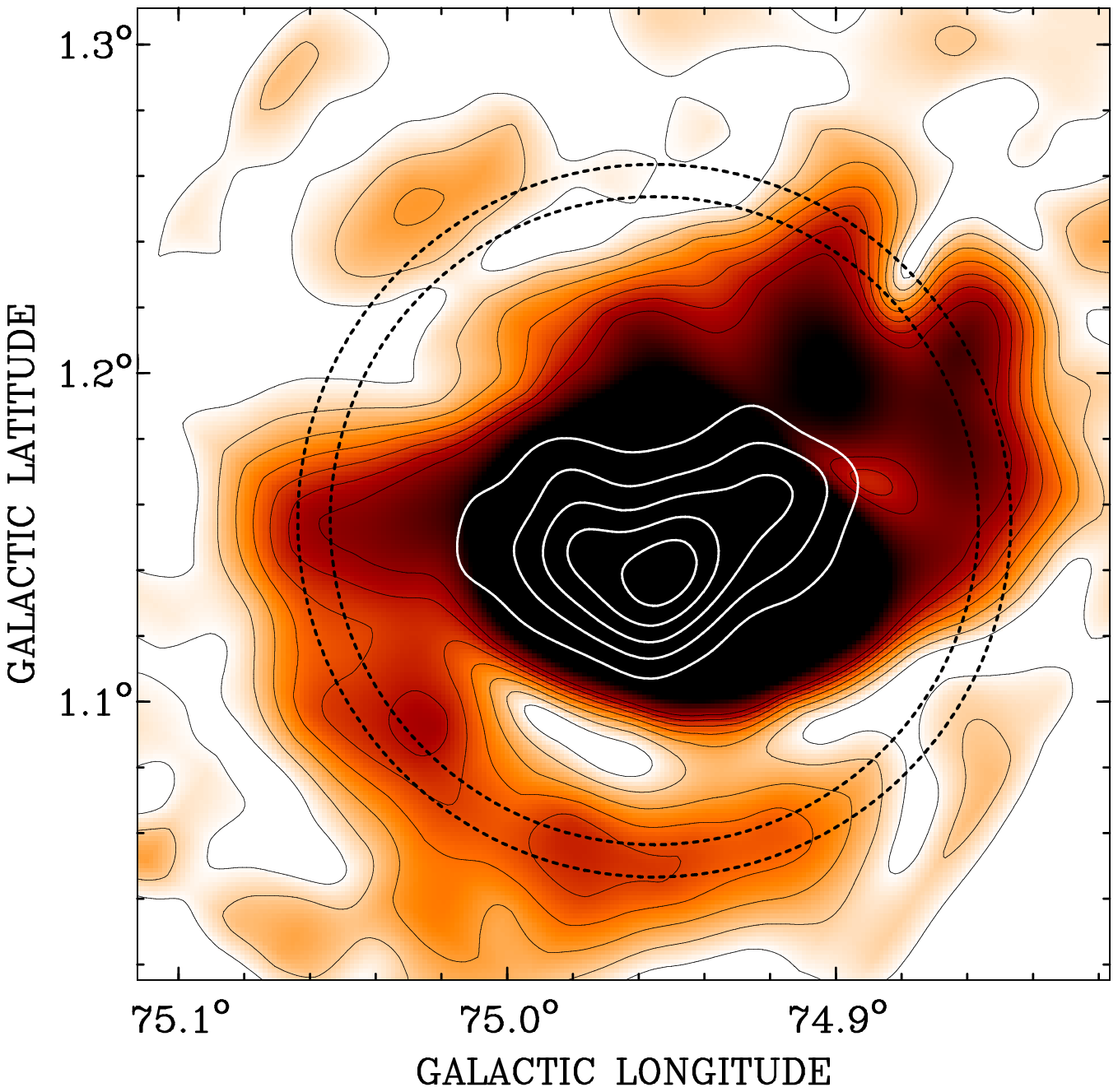}}
\caption{Saturated PI image at 4.75~GHz highlighting the radio shell candidate. Thick solid contours indicate the diffuse component (top panel, black and white contours) and the compact component (bottom panel, white contours). The two circles indicate the radius of the simulated shell inside the molecular cloud projected to the plane of the sky at 11.0 and 12.0~pc. The thin solid black contours outline the low level polarised emission at 4.75~GHz. Contours from 0 to 30~mK in steps of 5~mK ($\approx 2\sigma$) are displayed.}
\label{fig:shell}
\end{figure}

Our simulations of the two components of CTB\,87, the relic and the diffuse pulsar wind nebulae, revealed not only new results about the supernova explosion and the resulting pulsar but also give us an estimate on the radii of the blast wave inside the molecular cloud complex and the one expanding freely inside the large stellar wind bubble. For the shock wave, that entered the molecular cloud complex, we calculate a radius of $R_{CO}$ between 11 and 12~pc, projected to the plane of the sky. This radius is well confined since it mainly depends on the projected distance between the explosion of the supernova and the molecular cloud, which we know very well. For the shock wave expanding freely inside the stellar wind bubble or cavity a radius $R_{cav}$ between 30 and 35~pc is derived (see Table~\ref{tab:charac}). However, this value is not confined very well, due to the high age of the system and the uncertainty of the distribution of material inside the stellar wind bubble.

In Fig.~\ref{fig:shell}, we display the polarised intensity map at 4.75~GHz with contours indicating the diffuse (top) and relic (bottom) component. We plotted circles with radii of 11 and 12~pc indicating the possible location of the shock wave inside the molecular cloud complex, projected to the plane of the sky. The molecular cloud complex is to the south-east of CTB\,87. We find a candidate radio shell in our PI image that is not visible in total power at any frequencies. At 1420~MHz and 10.55~GHz in the polarisation images, there is also a marginal detection of this feature visible (see Fig.~\ref{fig:pi}). The radio shell in polarised intensity seems pretty wide, unusual for a typical limb-brightened shell of a supernova remnant, which are highly compressed, in particular in later stages of their evolution. Such a shell can appear to be wider if we are not looking entirely along this shell, but the shell is displaced along the line of sight and also has a velocity component towards us or away from us. This seems to favour the situation where the molecular cloud is displaced by an angle of $30\degr$ with the plane of the sky, the bottom case in Table~\ref{tab:charac}. The gap in polarised intensity between the polarised emission of the kidney-shaped feature and this polarisation shell indicates that this shell is an actual independent feature, in particular since the intrinsic magnetic field is the same (see Fig.~\ref{fig:rmmap}). If the intrinsic magnetic field was $90\degr$ apart, the gap could have been explained by beam depolarisation. But this is clearly not the case.



\section{Conclusions}

CTB\,87 is not a simple pulsar wind nebula but a very complex system of several different components that were produced in the same supernova explosion some 18,000~yr ago. 
The new radio polarimetric observations covering a wide frequency range from 1420 MHz to 32 GHz revealed that the CTB\,87 system contains two radio PWN components. One of those components is a compact radio relic with a strong magnetic field inside almost as high as DA\,495 \citep{koth08} and the Boomerang PWN \citep{koth06b}. It also has an unusually steep radio continuum spectrum, similar to other highly evolved PWNe, such as DA\,495 \citep{koth08} and G141.2+5.0 \citep{koth14}. It is the result of strong interaction of the supernova reverse shock with the wind of the central pulsar, with the closest approach between the reverse shock and the pulsar happened about 11,000 yr ago. The reverse shock was generated after part of the shock wave entered a molecular cloud complex to the south-east of the relic in the Galactic coordinate system. The remaining shock wave of the supernova explosion is expanding freely inside a large HI cavity or stellar wind bubble. 

Simulations of the evolution of the two PWN components within the system result in the likely explosion parameters of $E_0 \approx 7 \times 10^{50}$~erg with an ejecta mass of $M_0 \approx 12$~M$_\odot$ and intrinsic pulsar parameters of $\dot{E}_0 \approx 4 \times 10^{38}$~erg\,s$^{-1}$ and $\tau_0 \approx 2000$~yr.

The simulations also predict the current locations of the freely expanding shock wave inside the HI cavity or stellar wind bubble at 30 to 35~pc and the shock wave that entered the molecular cloud complex at 11 to 12~pc, projected to the plane of the sky, in a south-eastern direction in Galactic coordinates. We found a candidate radio shell towards the molecular cloud most pronounced in the polarisation data at 4.75~GHz. In total power, this shell is not evident. 

\section*{Acknowledgements}


The Dominion Radio Astrophysical Observatory is a National Facility
operated by the National Research Council Canada. The Canadian Galactic
Plane Survey is a Canadian project with international partners, and is
supported by the Natural Sciences and Engineering Research Council
(NSERC). SSH is supported by the NSERC Discovery Grants Program. The data are based on observations with the 100-m telescope of the 
MPIfR (Max-Planck-Institut f\"ur Radioastronomie) at Effelsberg.
We thank the referee for detailed comments that helped clarify the paper.

\bibliographystyle{mnras}
\bibliography{kothes} 












\bsp	
\label{lastpage}
\end{document}